\documentclass[pdflatex,sn-mathphys-ay]{sn-jnl}%

\usepackage{graphicx}%
\usepackage{multirow}%
\usepackage{amsmath,amssymb,amsfonts}%
\usepackage{amsthm}%
\usepackage{mathrsfs}%
\usepackage[title]{appendix}%
\usepackage[dvipsnames]{xcolor}%
\usepackage{textcomp}%
\usepackage{manyfoot}%
\usepackage{booktabs}%
\usepackage{algorithm}%
\usepackage{algorithmicx}%
\usepackage{algpseudocode}%
\usepackage{listings}%
\usepackage{xspace}

\newcommand\araa{\rmfamily{ARA\&A}}%
\newcommand\apj{\rmfamily{ApJ}}%
\newcommand\apjl{\rmfamily{ApJ}}%
\newcommand\apjs{\rmfamily{ApJS}}%
\newcommand\aap{\rmfamily{A\&A}}%
\newcommand\mnras{\rmfamily{MNRAS}}%
\newcommand\na{\rmfamily{New A}}%
\newcommand\pasa{\rmfamily{PASA}}%
\newcommand\rmxaa{\rmfamily{Rev. Mexicana Astron. Astrofis.}}%

\newcommand{\Lya}{\,\ifmmode{{\mathrm{Ly}\alpha}}\else Ly$\alpha$\fi\xspace}

\begin{document}

\title[Multi-phase gas]{Simulations of multi-phase gas in and around galaxies}

\author[1,2]{\fnm{Max} \sur{Gronke}}\email{max.gronke@uni-heidelberg.de}
\equalcont{These authors contributed equally to this work.}

\author[3]{\fnm{Evan} \sur{Schneider}}\email{eschneider@pitt.edu}
\equalcont{These authors contributed equally to this work.}

\affil[1]{
\orgname{Centre for Astronomy of Heidelberg University, Astronomisches Rechen-Institut 69120 Heidelberg}, \orgaddress{\street{Mönchhofstr. 12-14}, \postcode{69120} \city{Heidelberg}, \country{Germany}}}

\affil[2]{
\orgname{Max Planck Institute for Astrophysics}, \orgaddress{\street{Karl-Schwarzschild-Str. 1}, \postcode{85741} \city{Garching}, \country{Germany}}}

\affil[3]{\orgdiv{Department of Physics \& Astronomy and PITT-PACC}, \orgname{University of Pittsburgh}, \orgaddress{\street{100 Allen Hall, 3941 O'Hara Street}, \city{Pittsburgh}, \postcode{15260}, \state{PA}, \country{USA}}}

\abstract{Multiphase gas—ranging from cold molecular clouds ($\lesssim 100\,$K) to hot, diffuse plasma ($\gtrsim 10^6\,$K) is a defining feature of the interstellar, circumgalactic, intracluster, and intergalactic media. Accurately simulating its dynamics is critical to improving our understanding of galaxy formation and evolution, however, due to their multi-scale and multi-physics nature, multiphase systems are highly challenging to model. In this review, we provide a comprehensive overview of numerical simulations of multiphase gas in and around galaxies. We begin by outlining the environments where multiphase gas arises and the physical and computational challenges associated with its modeling.
Key quantities that characterize multiphase gas dynamics are discussed, followed by an in-depth look at idealized setups such as turbulent mixing layers, cloud-wind interactions, thermal instability, and turbulent boxes. The review then transitions to less idealized and/or larger-scale simulations, covering radiative supernovae bubbles, tall box simulations, isolated galaxy models including dwarf and Milky Way–mass systems, and cosmological zoom-in simulations, with a particular focus on simulations that enhance resolution in the halo. Throughout, we emphasize the importance of connecting scales, extracting robust diagnostics, and comparing simulations to observations. We conclude by outlining persistent challenges and promising directions for future work in simulating the multiphase Universe.}


\maketitle

\tableofcontents

\section{Introduction}\label{sec1}

In the classic theories of the interstellar medium (ISM), ``multiphase" refers to genuinely different phases of matter -- i.e. the cold and neutral clouds of molecular and atomic hydrogen, and the warmer, partially ionized volume-filling phase these clouds are embedded in \citep{Field1965, McKee1977}. Observationally, this is also commonly the definition that is meant. However, in more recent work on the circumgalactic medium (CGM) and intracluster medium (ICM), and particularly in simulations where the ionization state of gas is not necessarily tracked, ``multiphase" often refers to a two (or more) temperature plasma, where all the gas is ionized, but multiple temperature equilibria exist \citep[see also][]{Waters2023}. In this article, we apply a generic definition more similar to the latter case, where ``multiphase" simply refers to a system with gas at multiple temperatures that is roughly co-spatial.

A challenge then immediately arises when defining the different phases. In many of the physical systems covered herein, there are often two relatively stable (long-lived) phases, which are commonly referred to as `hot' and `cold'. However, there is also often more transient gas at intermediate temperatures generated by mixing or cooling (two major topics of this review). In addition, the actual temperature of the `hot' or `cold' phase can differ by orders of magnitude depending on the system being studied. In simulations of the the ISM, for example, `cold' typically refers to neutral or molecular gas with temperatures below $\sim 300$ K, while in studies of the CGM `cold' often means photoionized gas at $\sim 10^4$ K (which in the ISM-literature is commonly referred to as `warm'). Because these definitions often vary by subfield, we do not attempt to come up with a universal terminology in this review; rather we will follow the typical usage in the literature\footnote{Though even this is not always standardized.}, and clarify in each section what temperatures are meant. That said, unless otherwise noted, `hot' generally refers to low density gas with temperature $T \gtrsim 10^6$ K.

Because of the ubiquity of multiphase gas in astrophysical settings, it is naturally beyond the scope of this review to cover all the relevant (sub-)fields and numerical experiments. 
Noteworthy gaps include the very small scales where `particle-in-cell' simulations become relevant, and the very large scales of cosmological simulations and the intergalactic medium (for reviews of the former and the latter, see, e.g., \citealp{Marcowith2020} and \citealp{Naab2016, Gnedin2022}, respectively). 
Other systems with multiphase gas which are deliberately omitted include the Solar corona \citep{Antolin2022}, protostellar and protoplanetary disks, as well as (AGN) accretion disks.
Instead, we focus in this review on multiphase systems that influence galaxy formation and evolution more directly. Even with these omissions, the range of topics included in this review is quite broad. Other recent reviews that have covered topics we touch on here include \cite{Donahue2022} (halo-scale processes, including thermal instability), \cite{Faucher-Giguere2023a} (mixing layers, cloud crushing, etc.), \cite{Thompson2024} (galactic winds), and \cite{Ruszkowski2023} (cosmic rays). We hope the unique framing of this review, which focuses explicitly on simulation techniques and results while attempting to connect many different scales, will be a useful complement to these works.

\begin{figure}
\centering
\includegraphics[width=\linewidth]{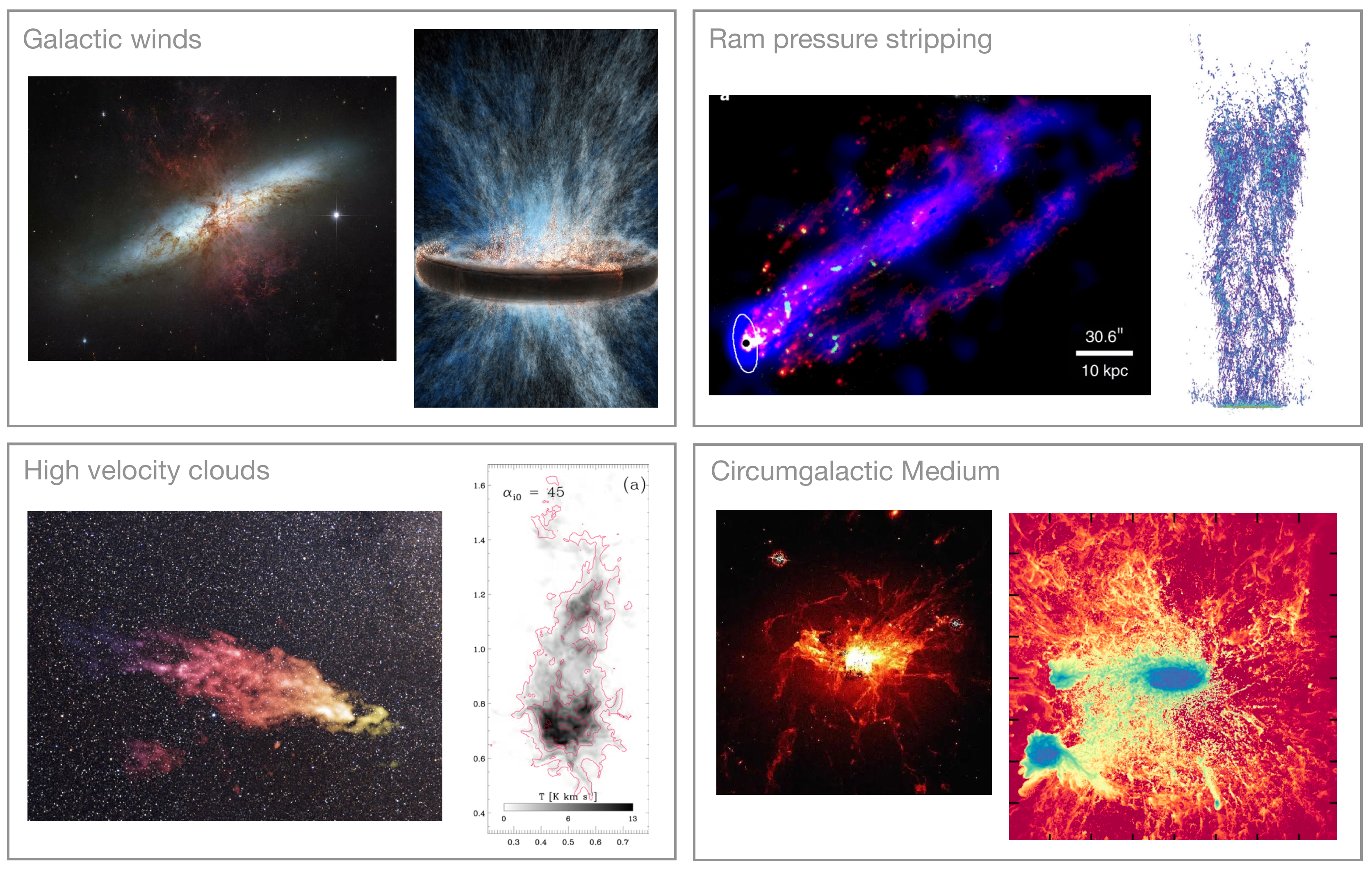}
\caption{Examples of observations and simulations of multiphase gas in and around galaxies. Sources/Credit: NASA/ESA/Hubble Heritage Project, \citet{Schneider2020} \textit{(top left)}; \citet{Tonnesen2012}, \citet{Sun2022} \textit{(top right}; NASA/ESA/Levay/STScI, Saxton/Lockman/NRAO/AUI/NSF/Mellinger,  \citet{Heitsch2016} \textit{(bottom left)}; \citet{fabian2008magnetic}, \citet{vandeVoort2019}.}
\label{fig:multiphase_examples}
\end{figure}

\subsection{Where multiphase gas exists}
\label{sec:where_multiphase_gas_exists}

When considering where multiphase gas arises in astrophysics, in some sense the answer is ``anywhere you look''. Some of the earliest theoretical studies of thermal instability were motivated by observations of solar prominences in the Sun’s chromosphere \citep{Thomas1952}. The ISM of our Galaxy is known to host gas ranging in temperature from a few Kelvin in the coldest regions of molecular clouds and planetary disks to tens of billions of Kelvin in the interior of supernova remnants — hence the common reference to “the multiphase ISM” \citep[e.g.][and references therein]{Draine2011}. Using observations across many different parts of the electromagnetic spectrum, similar conditions have been demonstrated to exist in the ISM of other local galaxies \citep[e.g.][the PHANGS-HST survey]{Lee2022}, and are now being catalogued for the most distant objects with JWST as well \citep[e.g.][]{Morishita2024}.

Conditions in the CGM of large galaxies are also ideal to maintain a multiphase medium, where the origin of the hot, volume-filling phase is generally considered to be a virial shock experienced by gas falling into a sufficiently large dark matter halo (see \S~\ref{sec:haloscale}), while the cooler phases can be seeded by thermal instability, ram pressure stripping of the ISM of satellite galaxies, or material ejected from the ISM of the central galaxy or satellite galaxies by winds. Regardless of its origin, observational evidence for multiphase gas in the CGM of massive galaxies, including our own, is ubiquitous \citep[see][for recent examples]{Peroux2019, Rudie2019, Fox2020, Lehner2020, Zahedy2021, Krishnarao2022, Sameer2024}. At the higher pressures found in the ICM, an even broader range of temperatures and densities is often observed \citep[e.g.][]{Rose2024}.

Not surprisingly, the transition region between the ISM and CGM of galaxies (sometimes called the disk-halo interface, though this region exists for other galaxy morphologies as well) is also a site of multiphase gas, where the generally hotter and lower density CGM material interacts with the warm and cool phases of the ISM \citep[see][for recent examples of observations in the Milky Way]{Bish2019, Vujeva2023, McClain2024}. One driver of this interaction is galaxy outflows, which are ubiquitous locations of multiphase gas, with observed temperature ranges over eight orders of magnitude that are essentially spatially coincident \citep[see e.g.][for examples across the spectrum of the outflow in the nearby starburst galaxy, M82]{Shopbell1998, Strickland2007, Westmoquette2009, Leroy2015, Lopez2020}. AGN winds have similarly been observed to host multiphase gas, which is often subject to both extreme radiation fields in the vicinity of accretion disks as well as extreme kinematics in the context of jets \citep[e.g.][]{Kosec2023, Zaidouni2024, Bischetti2024}.

Although the presence and interaction of multiphase gas is equally important on smaller scales (i.e. the solar corona, protoplanetary disks, and black hole accretion disks), and on the largest scales (i.e. the reionization and heating of the intergalactic medium), for brevity, this review will primarily focus on numerical experiments that seek to shed light on galaxy-scale phenomena. For recent reviews that focus on the observational picture of various systems discussed herein, see \citet[]{Tumlinson2017, Veilleux2020, Tacconi2020, Donahue2022, Hunter2024}, etc.

\begin{figure}
\centering
\includegraphics[width=0.55\linewidth]{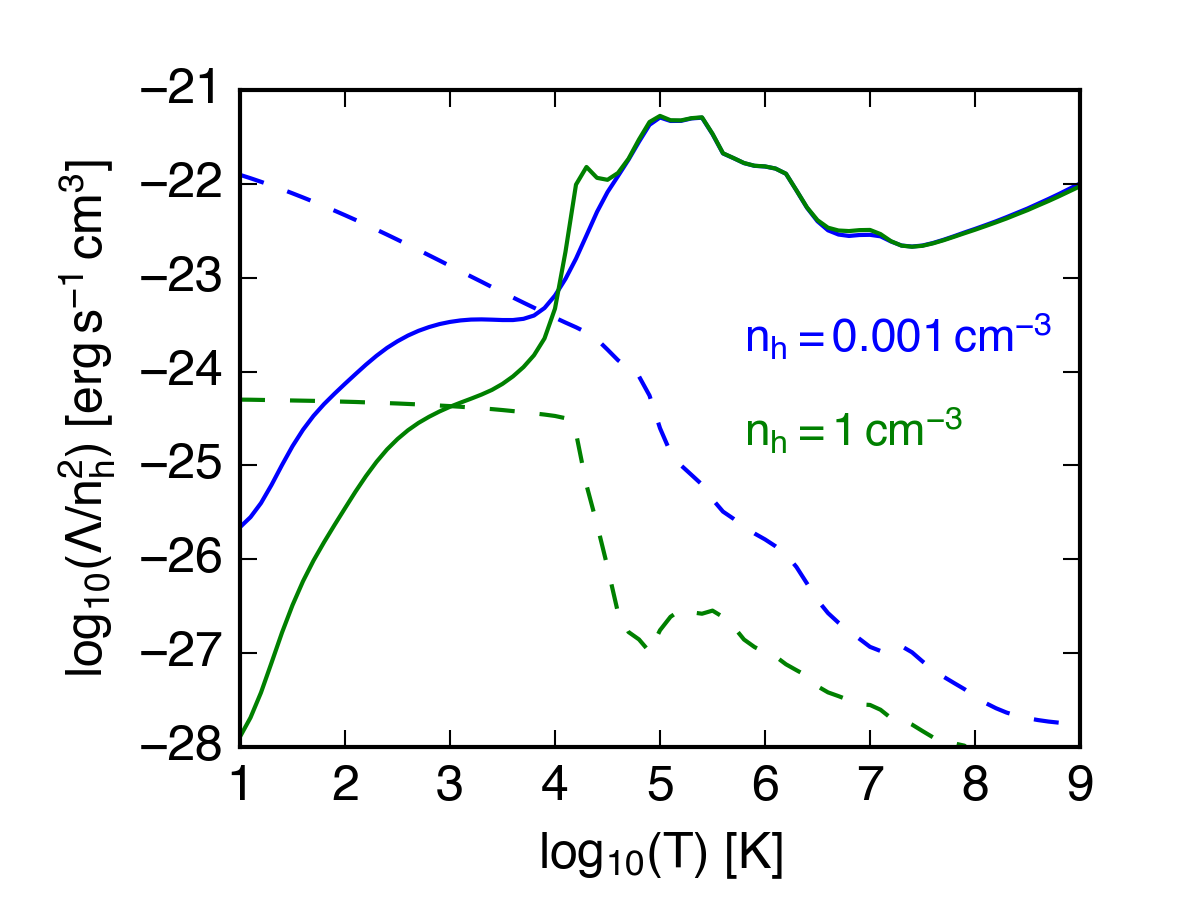}
\includegraphics[width=0.4\linewidth]{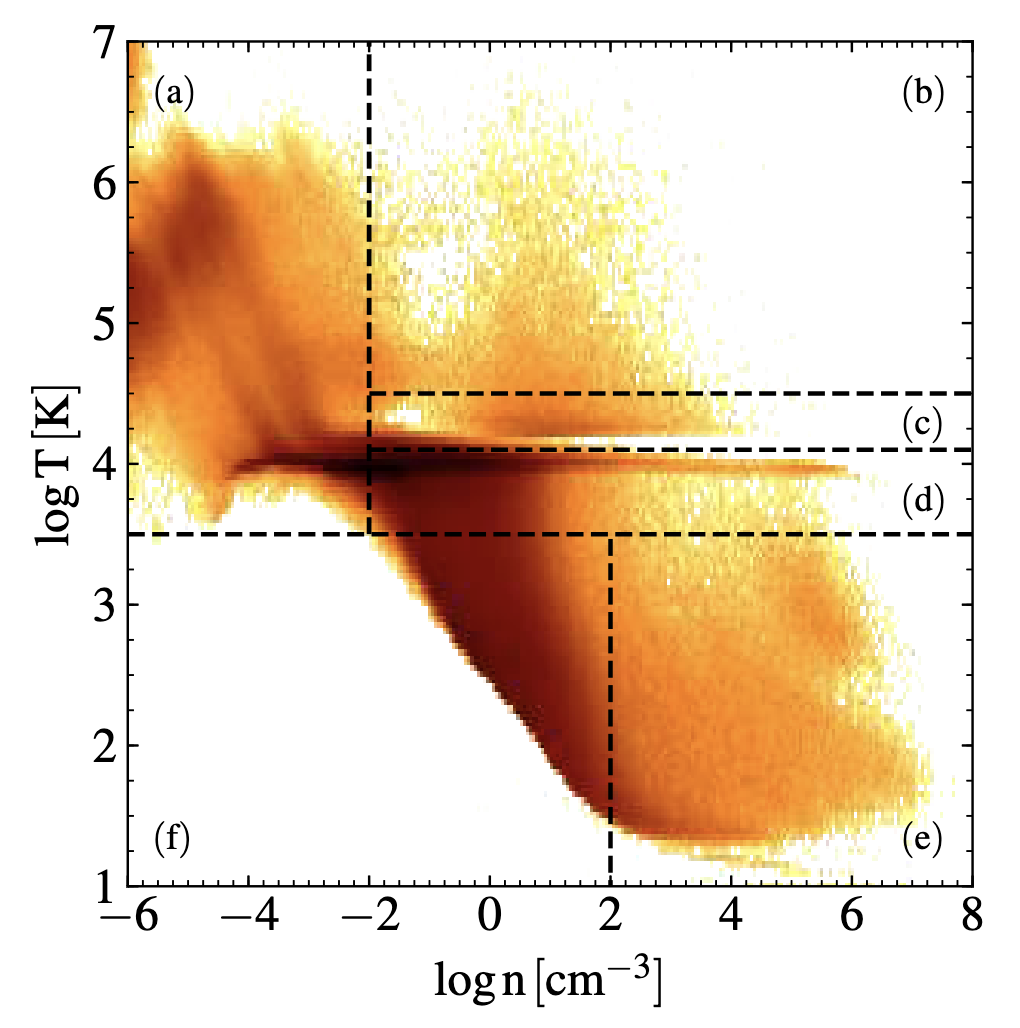}
\caption{Left: Cooling (solid lines) and heating (dotted lines) rates for solar metallicity gas at two different number densities subject to photoionization from the metagalactic UV background. Rates are generated using the Cloudy code and the `HM05' UV background table at redshift 0 \citep{Ferland2013}. Right (figure from \citealp{Marinacci2019}): Phase diagram from an isolated galaxy simulation showing many of the density-temperature regimes discussed in this review, including hot, low-density gas (upper left), ionized gas at $T\sim 10^4$~K, high-density cold gas in the interstellar medium, and intermediate temperature unstable gas between these regimes, all of which can have comparable pressure and interact in many multiphase systems, i.e. the ISM. The sharp boundary at the left side of the gas distribution reflects where heating and cooling are balanced, i.e. where the two curves cross for a given density in the left panel.}
\label{fig:coolingcurve}
\end{figure}

A classic explanation for the presence of multiphase gas is the action of thermal instability \citep[TI,][]{Field1965}. As originally suggested by \cite{Parker1953}, assuming that a gas is in thermal equilibrium where temperature independent energy gains are offset by temperature dependent radiative losses, TI will occur when radiative losses increase as temperature decreases. Such a condition is often satisfied in astrophysical systems, as when, for example, the slope of the radiative cooling curve is negative from $T \sim 10^5~-~10^7$ K (see Figure \ref{fig:coolingcurve}) and the primary source of heating is photoionization. Assuming a gas is initially in pressure equilibrium with a temperature near the inflection point at $T\sim 10^7$ K, slight density perturbations can lead to runaway cooling in overdense regions, as energy losses, $\dot{E}_\mathrm{cool}$, become more efficient with decreasing temperature. The re-establishment of pressure equilibrium will then result in compression of the cooling gas, which further increases the efficiency of energy losses (because $\dot{E}_\mathrm{cool} \propto n^2$).

The system will equilibrate when net heating again balances cooling in the overdense gas. Assuming conditions similar to those in the Milky Way halo with pressure $P/k_{B}\sim 10^2$ K $\mathrm{cm}^{-3}$, this will occur when cooling again becomes inefficient and is offset by photoionization heating from the UV background, typically around $T \sim 10^4$~K, though the exact temperature where the two rates balance depends on the density of the gas, the energy density of the ionizing photons\footnote{In the case of the UV background relevant in halos, the photon density depends on both the assumed model \citep[e.g.][]{Haardt2012, Puchwein2019, Faucher-Giguere2020} and the redshift. Within galaxies, this will typically depend on galaxy properties, such as the star formation rate, and distance to ionizing sources. For example, the equilibrium temperature is much higher in an HII region surrounding a young cluster than in the interior of a molecular cloud.}, and the gas metallicity
\footnote{This review primarily discusses simulations carried out assuming solar metallicity. While collisional cooling rates can depend sensitively on metallicity—particularly in the thermally unstable temperature range $T \sim 10^5$--$10^7\,\mathrm{K}$—many of the studies reviewed here focus on dimensionless behavior or on cooling time scales, which can be straightforwardly rescaled for different metallicities. As a result, the qualitative conclusions and physical interpretations are generally applicable beyond the solar-metallicity case.}
(see Figure \ref{fig:coolingcurve}). The gas can then maintain a stable equilibrium with two phases — a low density hot phase and a higher density cool phase, such that $n_\mathrm{cool} T_\mathrm{cool} = n_\mathrm{hot} T_\mathrm{hot}$, under the assumption of purely thermal pressure support\footnote{Additional forms of pressure support, such as magnetic fields or cosmic rays, can decrease the density of the cool phase such that this equality no longer holds.}. Because the density of the cooler gas will always be higher in this scenario, it will typically form small clumps that are distributed within the volume-filling hot medium \citep[see e.g.][Figure 4]{McCourt2018}.

Such an origin for multiphase gas is often referred to as ``precipitation”, and it can be driven by density perturbations in an initially static medium. However, equally important in many of the systems to be considered in this review are dynamical origins for multiphase gas — for example, mixing due to turbulence. In this case, gas which already contains multiple phases can be mixed together via hydrodynamic instabilities at interfaces to generate intermediate temperature and density gas, as for example in Kelvin-Helmholtz or Rayleigh-Taylor instabilities \citep{Chandrasekhar1961}. Shocks can also generate multiphase gas as they sweep through a medium and rapidly compress and heat gas, and thermal conduction can generate intermediate temperature gas at interfaces between hot and cold media.

In many real systems, of course, all of these mechanisms are at play at once. Although each of the aforementioned mechanisms has been studied extensively through analytic calculations (and much has been learned), this combined complexity (and the non-linearity of the long term evolution of many of these instabilities) is one of the primary drivers behind simulations of multiphase media.

\subsection{Computational challenges}
\label{sec:computational_challenges}

Modeling multiphase media is computationally challenging. As described above, typical temperature and associated density contrasts found in astrophysics range from $\sim 100$ to $\gtrsim 10^{10}$ making them numerically hard and the use of many adaptive techniques such as ``smoothed particle hydrodynamics" (SPH) problematic. In addition, the large temperature contrasts lead naturally to different length scales in the different temperature phases, making the problem inherently multi-scale. However, while multi-scale problems are common in astrophysics, multiphase environments such as the CGM or ICM have the additional complication that while cold gas may exist in very small-scale structures, it can be distributed over a large volume (see \S~\ref{sec:where_multiphase_gas_exists}). This fact makes traditional numerical techniques designed to deal with multi-scale problems such as adaptive mesh refinement (AMR, where, e.g., the cell sizes are adapted to the local density) inadequate or prohibitively expensive when attempting to resolve the multi-scale nature of multi-phase media. For instance, various observational probes have found cold gas structures in the CGM with scales as small as $\sim$ tens of pc \citep{Schaye2007,Lan2017,Tumlinson2017}. Thus, it would require\footnote{Assuming $\gtrsim 8$ cells per dimension are needed in order to resolve a structure.} $\gtrsim (8\times 100\,\mathrm{kpc} / 10\,\mathrm{pc})^3\approx 10^{15}$ resolution elements to model the multi-phase nature of a typical Milky-Way mass galactic halo\footnote{The number of resolution elements in the largest fixed-resolution hydrodynamic simulations to-date is $\sim10^{11}$ (see \S\ref{sec:tech_advances}).}. This leads to the fact that in typical cosmological simulations where the resolution is orders of magnitude worse, multi-phase quantities such as the amount of cold gas are unconverged \citep[e.g.][]{Faucher-Giguere2016}. This has been demonstrated explicitly in recent years in simulations in which the resolution of the CGM has been systematically enhanced (see ~\S~\ref{sec:ehanced_resolution_cgm}).

A final computational challenge in modeling multiphase media arises due to their multi-physics nature, that is, the dynamics of multiphase systems are often affected by several physical processes. Foremost are the effects of turbulence, which leads to the mixing (and destruction) of phases, and radiative cooling and heating, which leads to mass and momentum exchanges between the phases. Other non-negligible processes include thermal conduction and viscosity, which affect the temperature and turbulent properties of the gas, respectively, but also, e.g., the abundance of metals or molecules and the linked chemical processes, which directly affect the cooling -- and hence the multi-phase structure. Not to be forgotten are magnetic fields and cosmic rays, which also play a crucial role --  not only as non-thermal pressure sources but also by directly affecting the mixing processes. Naturally, this list is just ``the tip of the iceberg" as for instance the exact nature of heating (e.g., via cosmic rays or feedback processes) or turbulence (and the linked question of the drivers) can affect the evolution of a multiphase system.

For these reasons, to-date it is too computationally demanding to model all of the relevant scales and processes simultaneously and most simulations thus focus on specific physical questions and / or astrophysical systems. In \S~\ref{sec:simulations}, we list and discuss a wide range of simulations of astrophysical multiphase media, ordered approximately by scale and complexity.

\subsection{Key quantities of Multiphase Gas Dynamics}

When focusing on the dynamics of multiphase gas, most simulations in some way ask the question of what the thermal state of the medium is. Naturally, the evolution of the full phase-space diagram, i.e., the 2D distribution of gas in the temperature-density plane captures this information (see Figure \ref{fig:coolingcurve}, right panel). However, in practice it is often convenient to bin this diagram in terms of distinct temperature ranges (see ~\S~\ref{sec:where_multiphase_gas_exists}) leading to quantities describing the amount of e.g. mass and volume associated with a given phase. This leads to an additional challenge in comparing different works in the literature, as the exact definition of these temperature ranges may vary from one numerical study to another, and similarly, different subfields may disagree on the definition of e.g. `warm' versus `cool' gas. For the temporal evolution of particular interest are then the fluxes between the phases, e.g., in mass ($\dot m$) or momentum ($\dot{\boldsymbol{p}}$), and -- on the very coarsest level -- whether a certain phase exists or ceases to exist.

Another important aspect simulations try to capture is the kinematics of multiphase gas, i.e., how the different phases (co-)move with each other. Common ways of measuring this include simply the mean velocity difference between two phases ($\Delta v$) -- which can be mass weighted or evaluated at different points in space -- or again a full 2D distribution in temperature-velocity space. While these measures focus on the bulk flow of the phases, the turbulent kinematics can also be analyzed using the velocity-structure-function (VSF) or kinetic power spectra.

Beyond the amount of gas in different phases and its kinematics, understanding the spatial distribution and morphology of different phases is crucial for connecting simulations to observations. Higher-order quantities such as areal and volume filling fractions, clump size distributions, and phase coherence provide additional insights into the structure of multiphase gas. These measures not only help quantify how gas fragments and mixes but also enable a more direct comparison with observational constraints. Ultimately, radiative transfer tools such as SKIRT \citep{Camps2015} or \texttt{trident} \citep{Hummels2017} facilitate post-processing of simulations to generate synthetic spectra, allowing for a direct comparison with quasar sightline observations and other spectral diagnostics of multiphase gas.

A final factor to consider in studies of multiphase media is the characteristic spatial and temporal scales that must be resolved in order to obtain convergence in specific diagnostics. Unlike the case of gravitational collapse—where the \citet{Truelove1997ApJ...489L.179T} criterion provides a clear and widely adopted resolution requirement—there is no single universal criterion for multiphase flows. Rather, which scale must be resolved depends on the quantity of interest: achieving convergence in, for example, mass-transfer rates, phase morphology, or kinematics generally requires resolving the characteristic scale relevant to that process by at least a few grid cells. We will discuss some of these characteristic scales explicitly in \S~\ref{sec:simulations}, together with the particular diagnostics for which resolving them leads to numerical convergence.

\section{Simulations of Multiphase Media}
\label{sec:simulations}

\begin{figure}
\centering
\includegraphics[width=\linewidth]{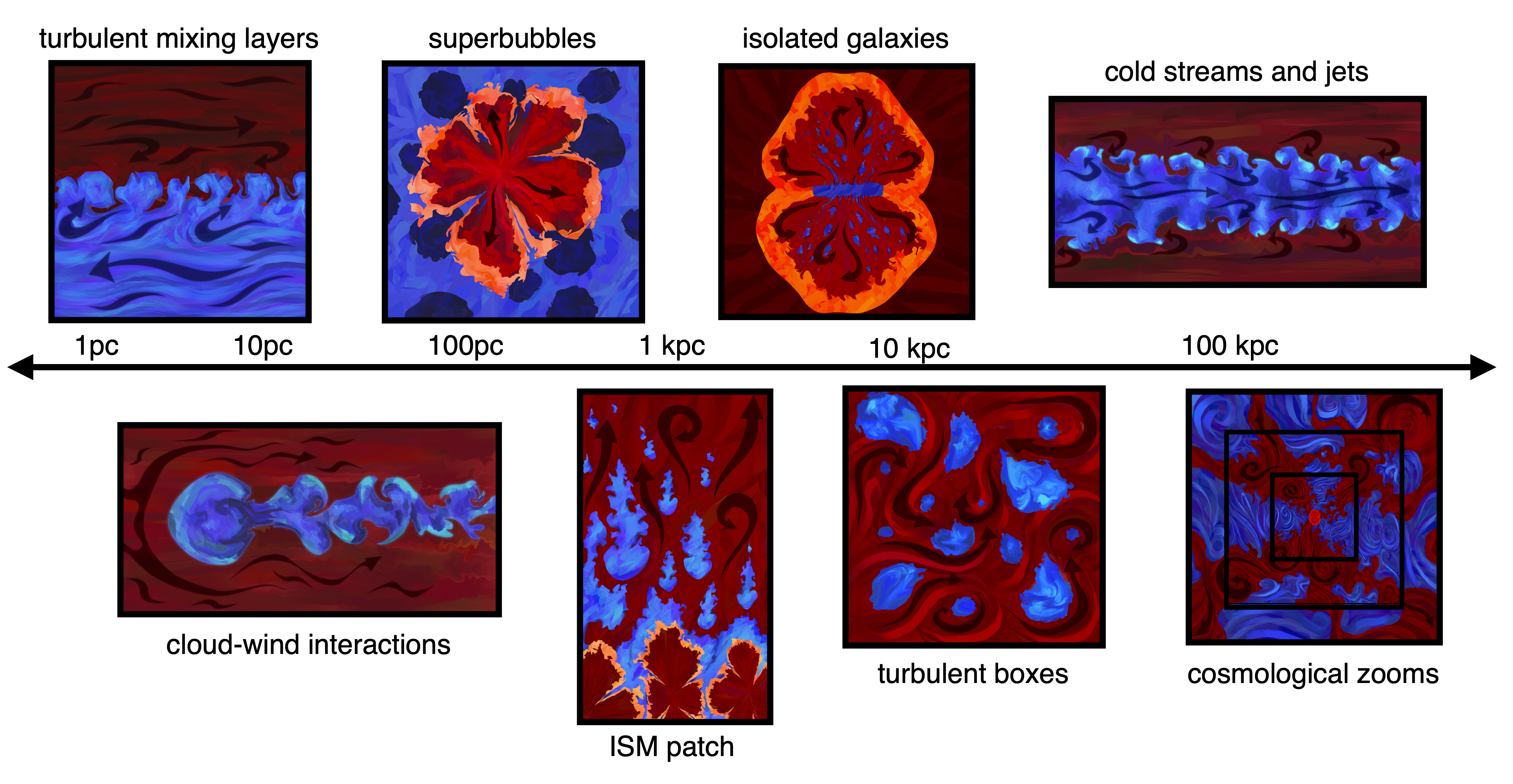}
\caption{Illustrations of the various simulation setups discussed in this review. Art by Giana Deskovich.}
\label{fig:simulations}
\end{figure}

In this section, we describe different simulations of multiphase gas flows found in astrophysics, organized approximately by increasing levels of physical scale and complexity. Figure~\ref{fig:simulations} shows a cartoon overview of the various setups discussed, along with their (roughly) associated length scales. We start with simple shear layers (\S~\ref{sec:tmls}), then describe cold clouds (\S~\ref{sec:cc}) and streams (\S~\ref{sec:streams}) embedded in a moving hot phase before moving on to turbulent multiphase media and thermal instability (\S~\ref{sec:turbulent}). We then shift to larger / less idealized setups, including supernova-driven bubbles in the ISM (\S~\ref{sec:superbubbles}), stratified simulations of patches of galaxy disks (\S~\ref{sec:ISMpatch}), larger isolated galaxy simulations (\S~\ref{sec:disks}), and halo-scale simulations (\S~\ref{sec:cosmo}).

\begin{figure}
    \centering
    \includegraphics[width=\linewidth]{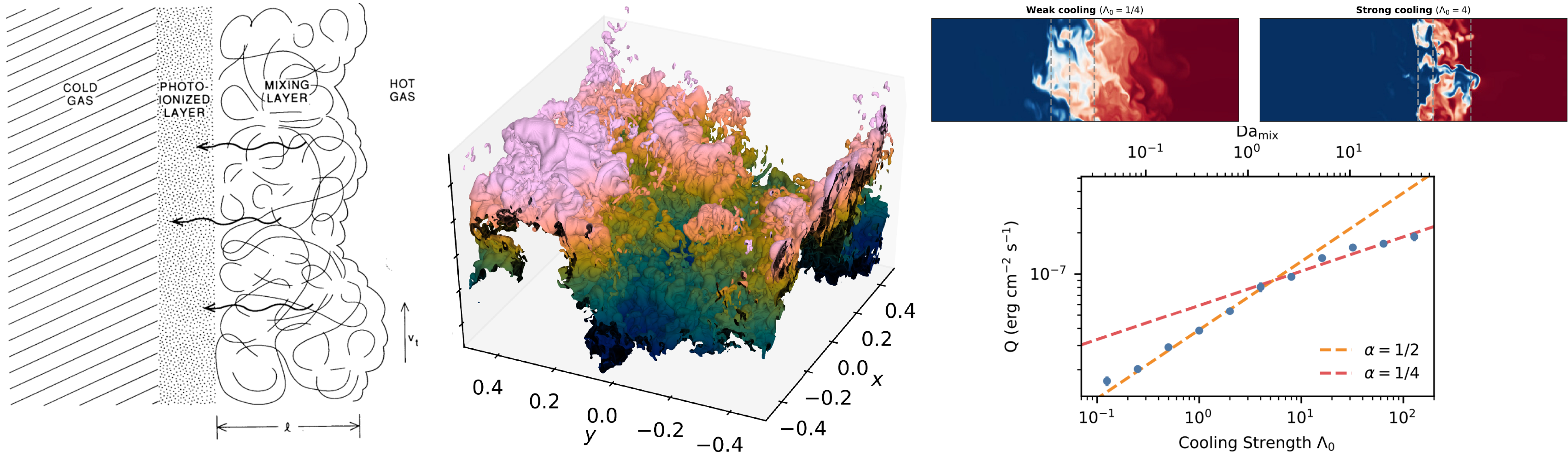}
    \caption{\textit{Left panel:} Schematic of a turbulent mixing layer from \citet{Slavin1993} with the flow moving vertically upwards. \textit{Central panel:} 3D rendering of the temperature isosurface of a turbulent mixing layer simulation \citep{Fielding2020}. \textit{Right panels:} Radiative, turbulent mixing layers shows two characteristic regimes both in morphology (top panels showing temperature slices) and radiated energy (bottom panel) separated by ${\rm Da}=1$ \citep{Tan2020}.}
    \label{fig:tmls}
\end{figure}

\subsection{Turbulent Mixing Layers}
\label{sec:tmls}

Shear-layer or turbulent mixing layer simulations are among the most fundamental and idealized setups for studying multiphase gas dynamics. In these simulations, 
two phases with densities $\rho_{\rm hot}$ and $\rho_{\rm cold}$ are placed on opposite sides of a (perturbed) interface and given a relative velocity $v_{\rm shear}$ parallel to their interface. This setup naturally leads to the development of the Kelvin-Helmholtz (KH) instability, which grows on a characteristic timescale $t_{\rm KH} \sim \chi^{1/2} \ell/v_{\rm shear}$ \citep{Chandrasekhar1961}, where $\chi\equiv \rho_{\rm cold}/\rho_{\rm hot}$ and $\ell$ is the perturbation length scale. This instability drives mixing and the exchange of mass, momentum, energy, etc. between the phases, making it a key process in the evolution of multiphase gas.

For a sharp interface (i.e., one that is smooth only on a scale $\ll \ell$, see below) -- as is commonly found in multiphase media where the cooling time of the cold phase is much shorter than other dynamical timescales -- the KH instability is unstable at all wavelengths permitted by the system.
Consequently, wherever there is shear between phases, mixing and mass exchange inevitably occur. 
However, several physical effects can modify or suppress this instability \citep{Landau1944, Chandrasekhar1961}, and (as we discuss further below) lead to reduced mixing and mass exchange between the phases. Among the most common processes causing suppression are:
\begin{itemize}
    \item \textit{Gravity:} The square root in the dispersion relation is negative, i.e., the instability occurs, if $\ell < v_{\rm shear}^2 / (g \chi)$
where $g$ is the acceleration pointing from the hot to the cold medium. This implies that for strong (effective) gravitational acceleration, the KH instability can be suppressed \citep[e.g.][]{Chandrasekhar1961,choudhuri1998physics}.

    \item \textit{Shear-layer thickness:} A smooth velocity gradient over a 
    range $\sim d$ suppresses instabilities at scales $\lesssim 10 d$ as modes smaller than this do not `see' the shear flow
    \citep{Chandrasekhar1961,Berlok2019a}. This implies that in viscous flows, the instability 
    is suppressed if the viscous timescale 
    $t_{\nu} \sim (\ell / 10)^2 / \nu$ (where $\nu$ is the kinematic viscosity) 
    is shorter than $t_{\rm KH}$ \citep{Roediger2013,Marin-Gilabert2022}.
    
    \item \textit{Supersonic flows:} For sufficiently high Mach numbers, 
    the KH instability is suppressed as the emergent perturbations are suppressed by subsequent incoming shocks. The critical threshold is given by \citep{Landau1944,Mandelker2016}
    \begin{equation}
        \mathcal{M}_{\rm crit} = (1 + \chi^{-1/3})^{3/2}
        \label{eq:Mcrit}
    \end{equation} where $\mathcal{M}\equiv v_{\rm shear}/c_{\rm s,hot}$.
    
    \item \textit{Magnetic fields:} A magnetic field component parallel to 
    the shear surface stabilizes the KH instability when the shear velocity 
    satisfies $v_{\rm shear} < (v_{\rm A,hot} v_{\rm A,cold})^{1/2}$, where 
    $v_{\rm A} = B_{\parallel}/\rho^{1/2}$ is the Alfvén velocity of the  respective medium with density $\rho$
    \citep{Chandrasekhar1961}. Even perpendicular magnetic fields can 
    suppress mixing by being bent into alignment with the shear, generating 
    a stabilizing parallel component over time \citep{Ryu2000,Das2023}.
\end{itemize}

The linear development and growth of the KH instability is well understood and often used as a benchmark for testing numerical codes \citep[e.g.][]{Lecoanet2017}. However, the nonlinear stage—where turbulence and secondary instabilities drive the full development of a mixing layer—is more complex, and requires the use of simulations to fully develop the theory.

Various studies have explored the role of different parameters on the growth of the mixing region. However, for multiphase gas, the inclusion of radiative cooling fundamentally changes the nature of the instability by enabling continuous mass exchange between the hot and cold phases. This radiative cooling aspect is particularly relevant for understanding astrophysical multiphase gas and is the focus of the following discussion.

\paragraph{Radiative, turbulent mixing layers:} Early numerical work on radiative turbulent mixing layers 
\citep{Esquivel2006,Kwak2010,Kwak2015,Ji2018} followed up on analytical work by \citet{Begelman1990} (extended by \citealp{Slavin1993}) who predicted the mean temperature of the mixing layer to be given by the flux (perpendicular to the interface) weighted average, i.e.,
\begin{equation}
  \label{eq:Tmean_BG90}
  \bar T = \frac{\dot m_{\mathrm{hot}} T_{\mathrm{hot}}+ \dot m_{\mathrm{cold}} T_{\mathrm{cold}}}{\dot m_{\mathrm{hot}} + \dot m_{\mathrm{cold}}}.
\end{equation}
\citet{Begelman1990} pointed out that while most gas is part of the hot or the cold phase, the intermediate temperature gas at $T=\bar T$ dominates the cooling, and thus, it is useful to consider turbulent mixing layers effectively as three phase media. They parametrized the hot gas mass flux into the mixing layer as
\begin{equation}
  \dot m_{\mathrm{hot}}\sim \eta_{\mathrm{hot}}\rho_{\mathrm{hot}} v_{\mathrm{mix}} A
\label{eq:mdothot}
\end{equation}
where $A$ is the surface area of the mixing layer parallel to the flow,
and $v_{\mathrm{mix}}$ the characteristic mixing velocity.
The efficiency factor $\eta_{\mathrm{hot}}$ arises because the entrainment of hot material into the layer is not perfect. This mass flux is linked to an enthalpy flux into the mixing layer given by $\dot e \sim 5/2 \eta_{\mathrm{hot}} P v_{\mathrm{mix}}$, where $P$ is the pressure. The main assumption of the \citet{Begelman1990} model is that this hot mass flow leads to turbulent eddy formation -- with scales $l_{\mathrm{cold}}<l_{\mathrm{t}}$ smaller than the total thickness of the layer $l_{\mathrm{t}}$ --which pulls the cold gas into the mixing layer at a rate
\begin{equation}
  \dot m_{\mathrm{cold}}\sim \eta_{\mathrm{cold}} \rho_{\mathrm{cold}} l_{\mathrm{cold}} / t_{\mathrm{KH}}
\end{equation}
where $\eta_{\mathrm{cold}}$ and $\rho_{\mathrm{cold}}$ is the corresponding mixing efficiency (see discussion below) and density of the cold medium, respectively. Crucially, the characteristic velocity $l_{\mathrm{cold}} / t_{\mathrm{KH}}$ includes the Kelvin-Helmholtz growth time $t_{\mathrm{KH}}\sim (\rho_{\mathrm{cold}}/\rho_{\mathrm{hot}})^{1/2} l_{\mathrm{cold}} / v_{\mathrm{mix}}$. Using these assumptions together with $\eta_{\mathrm{cold}}\sim \eta_{\mathrm{hot}}$, i.e., the efficiencies should be of similar magnitude (but see discussion below), in Eq.~\eqref{eq:Tmean_BG90}, one obtains as the characteristic temperature of the mixing layer the geometric mean of the hot and the cold temperatures $\bar T \sim (T_{\mathrm{cold}}T_{\mathrm{hot}})^{1/2}$
\footnote{Note that alternative analytical derivations based on geometrical or mass/enthalpy flux considerations come to the same conclusion (\citealp{Hillier2019}; \citealp[][\S~3.4]{Faucher-Giguere2023a})}.

It is important to note that the model does not predict the efficiencies $\eta$, thus even if $\eta_{\mathrm{hot}}\sim\eta_{\mathrm{cold}}$ holds, it is not predictive in terms of the influx rate Eq.~\eqref{eq:mdothot}, which is needed to understand the existence and growth of different phases. Furthermore, the model relies on rapid mixing and does not include other physical effects such as cooling or viscosity that might back-react on the development of turbulence \citep{Esquivel2006}. While different studies come to slightly different conclusions regarding the validity of the \citet{Begelman1990} model, they tend to agree (to within a factor of $\sim 5$) that it does accurately predict the intermediate ion abundances  \citep{Esquivel2006,Kwak2010,Kwak2015}, though \cite{Ji2018} potentially find larger deviations.\footnote{In these kinds of simulations it is important to reach steady state and have a sufficiently large box orthogonal to the flow direction. Some simulations of \citet{Esquivel2006} do not reach this steady state \citep{Kwak2010,Kwak2015}.} Note, however, that more recent work points out that obtaining the correct temperature distribution is required for obtaining the intermediate ion abundance which requires the inclusion of thermal conduction; we discuss this further below.

Later work also tries to qualitatively map and understand how the inflow into the mixing layer $\dot m_{\mathrm{hot}}$ as well as the overall structure and dynamics of the turbulent mixing layer depends on the geometry of the flow and other parameters.
Interestingly, it seems that the inflow of hot gas is of order the cold gas sound speed ($v_{\rm mix}\sim c_{\rm s,cold}$) and scales as $\dot m \propto t_{\rm cool}^{-1/2}$ for slow cooling \citep{Ji2018,Tan2020}, and $\dot m \propto t_{\mathrm{cool}}^{-1/4}$ for fast cooling (\citealp{Gronke2019,Tan2020,Fielding2020}) where `slow' and `fast' are with respect to the speed of the mixing  -- with the characteristic mixing timescale $t_{\rm mix}\sim \ell / u'$. Here, $u'$ denotes the local turbulent velocity. Specifically, the two regimes are divided by the corresponding `Damk\"ohler number' $\mathrm{Da}\equiv t_{\rm mix}/t_{\mathrm{react}} = t_{\mathrm{mix}}/t_{\mathrm{cool}}$ of unity (see Figure~\ref{fig:tmls}). The analogy of turbulent radiative mixing layers to turbulent combustion fronts -- where things burn (on a reaction timescale $t_{\mathrm{react}}$) versus cool (on a timescale $t_{\mathrm{cool}}$) was pointed out by \citet{Tan2020}.

Another interesting aspect of mixing layers is the fractal nature of their interface \citep{Fielding2020,Tan2020}. 
Notably, \citet{Fielding2020} demonstrated this behavior and developed an analytical framework that explicitly builds on the fractal morphology of the mixing layer seen in simulations. While the scalings found differ slightly from the combustion inspired model of \citet{Tan2020}, importantly, these numerical results show that indeed mixing and subsequent cooling is the important physical process 
 operating under conditions of approximate pressure equilibrium 
\citep{Fielding2020,Chen2022}\footnote{\citet{Fielding2020} demonstrated that not resolving $\mathrm{min}(c_{\rm s}t_{\rm cool})$ (cf. \S~\ref{sec:turbulent} for a discussion of this `shattering' length) leads to numerical pressure dips across the layer.}.

Since mixing on large scales takes longer (i.e., the eddy turnover time increases with eddy size; specifically $t_{\rm eddy} \propto \ell^{2/3}$ in a Kolmogorov cascade), the larger scales are the bottleneck of the mixing/cooling process.
This implies that while for instance the morphology of the mixing layers is generally not converged, the mass transfer rates are \citep{Tan2020}. Furthermore, because turbulent diffusion dominates, thermal conduction is mostly unimportant for the overall mass transfer or total luminosity of the layer \citep{Esquivel2006,Tan2020}.
However, for the full temperature distribution or the emission of the different intermediate temperate ions, thermal conduction plays a big role \citep{Tan2021}. This means that for larger scale simulations `only' the driving scale of a given mixing layer has to be resolved in order to converge in the mass transfer rate -- but for comparison with observations of, e.g., OVI the \citet{Field1965} length should be (or this has to be added in post-processing)\footnote{With classical \citet{Spitzer1962pfig.book.....S} conduction, $\lambda_{\rm F}\propto T^{5/2}$, so it might become unresolved below a cutoff temperature. \citet{Tan2021} showed that the temperature PDF converges only above this cutoff.}.

While the above studies mostly focused on the sub- and transonic regime, \citet{Yang2022} expanded this picture to higher Mach numbers (focusing on the slow cooling $\mathrm{Da}<1$ parameter space). They found that the turbulent velocity saturates at $\mathcal{M}\gtrsim 1$ and the mass flux is reversed: cold gas evaporates. How this behavior changes with more efficient cooling has yet to be explored.

\citet{Ji2018} (again focusing on the $\mathrm{Da}<1$ regime) and more recently \citet{Das2023} studied the effects of magnetic fields on turbulent radiative mixing layers. They showed that $B$-fields are amplified to equipartition and depending on the orientation can suppress the mass flux from the hot to cold medium significantly. This is in particular the case in the fast cooling ($\mathrm{Da}>1$) regime where mixing is the bottleneck.

\subsection{Cloud-wind interactions}
\label{sec:cc}

\paragraph{Basic considerations}
`Cloud crushing' simulations are central to understanding multiphase gas flows. In the most basic version, the setup is simple: a spherical cold cloud with density $\rho_{\mathrm{cold}}$ and radius $r_{\mathrm{cl}}$ is exposed to a wind with density $\rho_{\mathrm{hot}}$ and velocity $v_{\mathrm{wind}}$. This setup is partially motivated by the classic ISM picture of \citet{McKee1977} and follow up work \citep{Mckee1978} in which the question of how cold gas clouds interact with hot supernovae bubbles is posed -- and already raises the issue of how cold `high-velocity' clouds can be accelerated. This question of cold gas acceleration (in the context of galactic winds) remains an active topic of research today (for reviews of galactic winds including the question of entrainment, see \citealp{Veilleux2005,Zhang2018,Veilleux2020,Thompson2024}). This `cloud crushing' setup and variations thereof offers a rich physics problem and has been studied in $>50$ papers to date\footnote{See, e.g., a list of `cloud crushing' studies and their parameters at \url{http://bit.ly/cloud-crushing} (courtesy of W. Banda-Barr\'agan).}.

That this multiphase gas interaction leads to the destruction of the cold cloud was clear early on \citep[e.g.][]{Nittmann1982}, but \citet{Klein1994} established in a seminal paper that the shear and acceleration due to ram pressure lead to Kelvin-Helmholtz and Rayleigh-Taylor instabilities forming as well as a shock crossing through the cloud (see also \citealp{Murray1993}). These three processes destroy the cold gas on a `cloud crushing' timescale
\begin{equation}
  \label{eq:tcc}
  t_{\mathrm{cc}}\sim \chi^{1/2} \frac{r_{\mathrm{cl}}}{v_{\mathrm{wind}}}.
\end{equation}
(equal to the characteristic growth times of the instabilities, cf. \S~\ref{sec:tmls})
where again $\chi\equiv \rho_{\rm cold}/ \rho_{\rm hot}$ is the cold gas overdensity.

This timescale can be contrasted to the time it would take the hot wind to impart enough momentum to the cloud to accelerate it
\begin{equation}
  \label{eq:tdrag}
  t_{\mathrm{drag}}\sim \chi \frac{r_{\mathrm{cl}}}{v_{\mathrm{wind}}}.
\end{equation}
The fact that $t_{\rm drag}/t_{\rm cc}\sim \chi^{1/2}\gg 1$, i.e., the cold gas will be destroyed before it can be accelerated is precisely what gives rise to the puzzle of how cold gas can be accelerated to observed velocities of hundreds to even thousand of kilometers per second \citep{Veilleux2020} and is also known as the `entrainment problem' (see \citealp{Zhang2015a} for a detailed discussion).

Several papers focused on this cold-hot interaction and in particular the mixing and destruction process in high-resolution three dimensional simulations \citep[e.g.][]{Cooper2009,Schneider2015,Bruggen2016}. They also expanded the parameter space to higher overdensities \citep[e.g.][]{Cooper2009}, Mach numbers \citep[e.g.][]{Scannapieco2015a}, different cloud morphologies \citep[e.g.][]{Banda-Barragan2018}, and additional physical processes \citep[e.g.][]{Bruggen2016}.

For instance, the inclusion of magnetic fields was shown to change the morphology of the cold gas \citep{MacLow1994,Dai1994,Shin2008,Gronnow2018}, suppress the mixing, and lead to faster acceleration \citep{Dursi2008}. In fact, \citet{McCourt2015} claimed that increased magnetic drag could solve the `entrainment problem'.
 This was later debated by \citet{Gronke2020}, who showed that magnetic fields can only decrease the acceleration time by a factor of a few, making this solution feasible only for low ($\chi\lesssim 30$) overdensities, while for most astrophysically relevant overdensities of $\gtrsim 100$ one would require very strong (plasma beta $\beta < 1$) magnetic fields for the effect to allow cold gas survival.

\paragraph{The effect of radiative cooling}
An particularly important additional physical effect is the inclusion of radiative cooling. This was already alluded to in the classic \citet{Klein1994} paper (see their section 2.3.1) and studied further in later work \citep[e.g.,][]{Mellema2002}. However, while several detailed, high-resolution simulations included (strong) cooling \citep[e.g.,][]{Scannapieco2015a,Bruggen2016,Schneider2016}, most still concluded that eventually all cold gas is mixed into the hot wind (underlying the `entrainment problem'). This was primarily a result of the simulations either not following the evolution of the cloud long enough, not having large enough domains or not employing cloud tracking methods to capture the tail evolution, or only modeling clouds that were small enough that they were always destroyed. However, simulations by \citet{Marinacci2010} (in 2D) and later \citet{Armillotta2016,Armillotta2017,Gritton2017} (also in 3D) showed that for some clouds, the inclusion of radiative cooling halts the disruption and that in fact  the cold cloud mass can even increase.

\citet{Gronke2018} formulated a quantitative criterion for cloud survival and growth, namely that clouds can survive to be entrained if they fulfill the requirement
\begin{equation}
  t_{\rm cool,mix}/t_{\rm cc} < 1
  \label{eq:tcoolmixtcc}
\end{equation}
where $t_{\rm cool,mix}$ is the cooling time of the mixed gas defined -- in the spirit of \citet{Begelman1990} (see \S~\ref{sec:tmls}) -- using the geometric mean temperature and density between the hot and the cold medium.
Note that because the destruction time $t_{\rm cc}\propto r_{\rm cl}$ but the cooling time is merely depending on the properties of the gas, the survival criterion Eq.~\eqref{eq:tcoolmixtcc} can be recast to a geometrical threshold $r_{\rm cl}>r_{\rm crit}$, i.e., only clouds larger than a critical size can survive \citep{Gronke2018}. For typical wind / ISM conditions $r_{\rm crit}\sim \mathcal{O}(1)\,$pc\footnote{The survival criterion Eq.~\eqref{eq:tcoolmixtcc} converts to $r_{\rm crit}=2 \, {\rm pc} \ \frac{T_{\rm cl,4}^{5/2} \mathcal{M}_{\mathrm{wind}}}{P_{3} \Lambda_{\rm mix,-21.4}} \frac{\chi}{100}$ where $T_{\rm cl,4} \equiv (T_{\rm cl}/10^{4} \, {\rm K})$, $P_{3} \equiv nT/(10^{3} \, {\rm cm^{-3} \, K})$, $\Lambda_{\rm mix,-21.4} \equiv \Lambda(T_{\rm mix})/(10^{-21.4} \, {\rm erg \, cm^{3} \, s^{-1}})$, $\mathcal{M}_{\mathrm{wind}}$ is the Mach number of the wind with respect to the winds sound speed.} but this depends on the exact wind properties\footnote{Note because $r_{\rm crit}\propto 1/n$, the survival threshold can be rewritten as a fixed cloud column density.}.

The exact form of this survival criterion is still being debated. \citet{Li2019a} suggested the use of $t_{\rm cool,hot}/t_{\rm life} < 1$, where $t_{\rm cool,hot}$ is the cooling time of the hot medium, and $t_{\rm life}$ is an empirical function of the cloud disruption time (depending on several parameters but generally $\sim 10\,t_{\rm cc}$), which was later supported by \citet{Sparre2020} who compared the \citet{Gronke2018} and the \citet{Li2019a} criteria. A similar comparative study was conducted by \citet{Kanjilal2020} who came to the reverse conclusion.

Although these discrepancies seem puzzling, it is important to note that different groups adopt different criteria for what constitutes `survival' of the cold gas. This is particularly relevant because, in the initial 
phase ($t \sim \chi^{1/4}\,t_{\rm cc}$), the cold gas mass can drop substantially before increasing again later on. Thus, employing an arbitrary cutoff (e.g., defining `survival' as the cold gas mass not falling below $10\%$ of its initial value) though it can help avoid misinterpreting underresolved later evolution of the cloud, might not adequately capture the long-term evolution of the cold gas, and clouds that may eventually grow can potentially be flagged as `destroyed'.

Most recently, several studies pointed out that both of the criteria discussed above are somewhat unphysical and should be modified \citep{Abruzzo2021,Abruzzo2023,Farber2021}. First, as discussed in \S~\ref{sec:tmls} the `mixed' gas possesses a range of temperatures and overall the \citet{Begelman1990} picture of mixing layers is too simplistic, and second, the dominant process of temperature reduction at temperatures close to the hot medium is mixing and not radiative cooling. This can be demonstrated explicitly by switching off cooling above some temperature (in which case, clouds still grow). In reality, the actual effective cooling temperature entering Eq.~\eqref{eq:tcoolmixtcc} should be a weighted mean of $t_{\rm cool}$ between $T_{\rm cold}$ and $T_{\rm hot}$ \citep[see][]{Farber2021,Abruzzo2022}. However, while it is clear from the aforementioned studies (as well as from turbulent mixing layer simulations; \S~\ref{sec:tmls}) that the weighting function peaks around the minimum of the cooling time, its actual form is still unknown.

While a better theoretically motivated, generally valid `survival criterion' is still outstanding, some key points have been learned from the recent developments in radiative cloud crushing simulations:
\begin{itemize}
\item it is possible to have cold gas survive ram pressure acceleration if the cooling is efficient enough, i.e., some cooling time is shorter than some destruction time. This survival criterion $t_{\rm cool}<t_{\rm destroy}$ can be rephrased to a geometrical criterion $r_{\rm cl} > r_{\rm crit}$ which generally yields that clouds larger than $\sim $several parsec  will survive (in high pressure ISM/wind conditions; the threshold strongly depends on the environmental properties; see \citealp{Gronke2018}).
\item in the long-term evolution, clouds that survive continuously grow, leading to (potential) significant mass growth during their later evolution. This growth continues even as the cold gas is entrained and is driven by pulsations of the cold gas due to `overshooting' of the gas pressure \citep{Gronke2020,Abruzzo2022}. The mass transfer follows the scaling expected from TML simulations, i.e., $\dot m \propto t_{\rm cool}^{-1/4}$.
\item importantly, this cold gas growth happens predominantly in the wake of the cloud, making the required computational domain size large (the tails grow to $\sim 0.2-0.5 \chi r_{\rm cl}$) and, hence, `cloud tracking' techniques which continuously change the reference frame (as employed, e.g., in \citealp{Shin2008,McCourt2015,Bruggen2016} and described in detail in \citealp{Dutta2019}) are important.
\item the mass transfer from the hot to the cold medium also implies a momentum transfer leading to faster acceleration and entrainment of the cold cloud \citep{Gronke2020,Tonnesen2021}. This leads to efficient acceleration even for high density contrast ($\chi \sim 10^4$) molecular clouds \citep{Chen2024}.
\item compared to early studies -- which were mostly conducted in 2D, with short boxes, without radiative cooling, and only for a short dynamical timescale -- the above listed results were only made visible and possible because of computational advances. This allowed in particular following turbulent mixing and cooling (which requires sufficient resolution, 3D, and radiative cooling), as well as seeing the long-term growth of the cold gas (which requires long boxes and simulated timescales). 
\end{itemize}

\begin{figure}
    \centering
    \includegraphics[width=\linewidth]{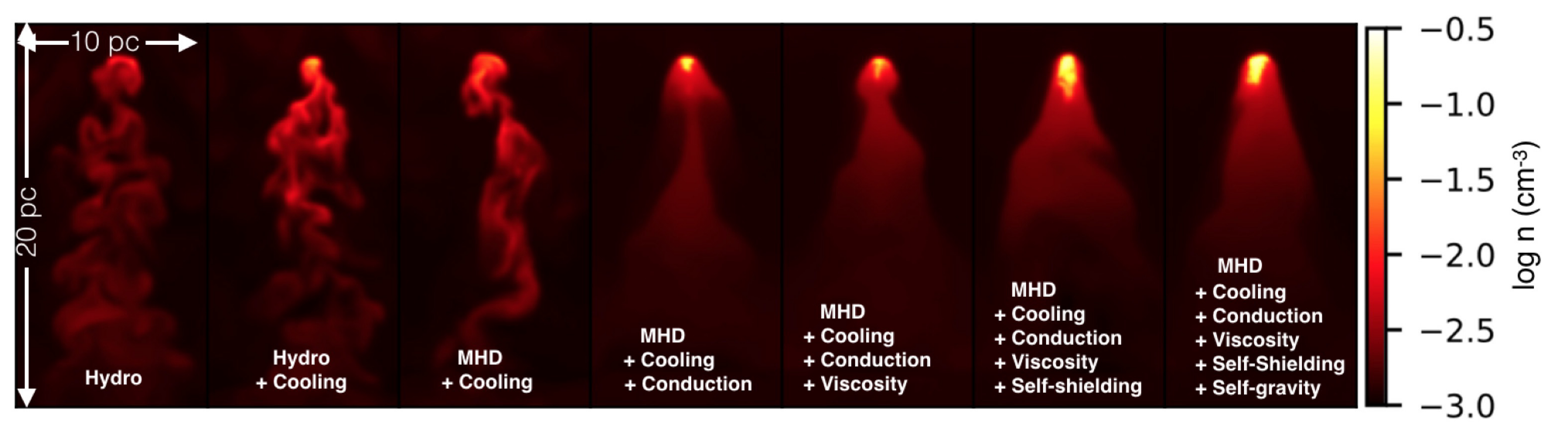}
    \caption{Different physical mechanisms affecting `cloud-wind' simulations (figure adapted from \citealp{Li2019a}). The panels show density slices of the `destruction regime' (i.e. with no or weak cooling) with the hot wind coming from the top.}
    \label{fig:cc_li}
\end{figure}

\paragraph{Additional physical effects}
An interesting recent development in the field has been to attempt to connect several of the physical effects discussed above. For instance, \citet{Hidalgo-Pineda2023} confirmed that magnetic fields alone cannot support the entrainment of $\chi\gtrsim 100$ clouds, and that the combination of magnetic fields with radiative cooling changes the entrainment process significantly, such that the required survival criterion shifts by several orders of magnitude to $t_{\rm cool,mix}/t_{\mathrm{cc}}< 100$ at a plasma beta of $\beta\sim 1$ in the wind. Similarly, \citet{Bruggen2023} qualitatively found that anisotropic conduction in combination with radiative cooling can aid cloud survival in the case of parallel magnetic fields but has a weak effect in the case of transverse magnetic fields (also see \citealp{Li2019a} who performed a large parameter scan; cf. Figure~\ref{fig:cc_li}).

Although some earlier simulations included lower temperature ($T<10^4\,$K) gas \citep[e.g.][]{Schneider2016}, the development of the above discussed survival criteria has renewed interest in this low temperature regime.
This is particularly important for extending the applicability of the `survival criterion' to colder gas, as well as for incorporating dust and molecular physics into the simulations \citep{Girichidis2021,Farber2023,Chen2024,Richie2024}. The latter is especially relevant for distinguishing between `in-situ' formation of cold gas in the halo and its transport from the ISM into the CGM. 

One important future direction will be to connect these idealized `cloud crushing' simulations to more realistic initial and boundary conditions. Several studies have already considered the impact of non-spherical, `turbulent' clouds \citep[e.g.][]{Schneider2016,Banda-Barragan2018,Banda-Barragan2019,Gronke2020} and found that in combination with cooling (i.e., for clouds that fall within the survival criterion), the longer term effects are small. More recently, studies like \citet{Banda-Barragan2020a} have increased the complexity of the cold gas morphology to further bridge the gap between `cloud crushing' and full ISM simulations (see \S~\ref{sec:ISMpatch}) -- a direction where more work is needed in order to generalize the `spherical cloud' survival criterion. One important physical effect not present in simulations of individual clouds are combined effects such as `shielding' or the change in the mixed gas properties because of multiple sites/radii at which a hot wind engulfs cold clumps, which has been studied by several groups \citep{Melioli2005,Pittard2005,Aluzas2012,Aluzas2014,Forbes2019}; still, more work is needed to connect these multi-cloud studies to the recent developments in understanding single-cloud evolution discussed above.

Curiously, while many studies have altered the cold gas properties, the effect of changing the wind conditions has been focused on less. However, a constant, homogeneous wind is not realistic for galactic winds (see \S \ref{sec:disks}) and should be considered in particular for the longer-term evolution ($t\gtrsim t_{\rm drag}$) of the cold cloud material. Critically, the adiabatic expansion of the wind leads to a drop in pressure and density, and thus, to an increase in the wind's Mach number \citep{Chevalier1985}. \citet{Gronke2020} attempted to include this in their boundary conditions (also see \citealp{Dutta2019}) and found that while the drop in wind density would lead to a slowdown of the mass transfer rate from hot to cold medium, this is mostly compensated by the increase in cold gas surface area due to the decrease in pressure. 

In fact, in a influential paper \citet{Fielding2022} incorporated the transfer terms between the two phases into a semi-analytical two-fluid wind model (see also \citet{Nikolis2024} who compared this model explicitly to cloud-crushing simulations) and showed that this change in background not only changes the cloud properties but that the back-reaction of the cold gas onto the hot medium can change the larger scale impact of the wind significantly.

Hence, more work is certainly needed to account for more realistic wind profiles and properties in cloud-wind simulations, which do not typically follow the idealized adiabatic expansion model (see \S~\ref{sec:disks}).

\paragraph{Infalling clouds}
While the above discussion focuses mostly focused on outflowing clouds, a similar `entrainment problem' also applies for infalling cold gas -- such as in coronal rain, gas infalling in the ICM, or the observed `high-velocity clouds' in the halo of the Milky Way \citep{Putman2012,Antolin2022} -- and thus the results above are sometimes discussed in the context of a static hot and a moving cold medium. However, while in some evolutionary stages the one reference frame can be transformed into the other, an important difference is that while an outflowing cloud is being entrained the shear generally decreases; this is not the case for an infalling cloud where the shear reaches a constant terminal velocity and, thus, is never `safe' from destruction. Only some studies have attempted to do so. While it is possible to simply simulate a `global box' \citep[as done, e.g., in][]{Martinez-Gomez2020} this is generally computationally prohibitive if one wants to focus on the detailed mixing processes. Instead, it is advantageous to modify the boundary conditions of a computational domain comoving with the cloud \citep[as done, e.g., in][]{Heitsch2009}. Studies conducted in this manner can follow a falling HVC in great detail and in fact are often used to produce mock observations; e.g., \citet{Heitsch2021} showed that mixing and accretion of halo gas can dominate the mass of an HVC and result in interesting metallicity and velocity gradients across the (stretched out) cloud, which can potentially reveal the accretion history of a cloud.

In a similar setup, \citet{Tan2023} focused on the dynamics of infalling clouds. They showed that due to the near constant shear at late times the survival criterion becomes more stringent (they argue for comparison of the mass doubling with the destruction timescale, i.e., $t_{\rm grow}\equiv m/\dot m \lesssim t_{\rm cc}$). Furthermore, they quantified that the mass accretion of halo gas onto the HVC leads to a terminal velocity $v_{\rm t,grow}\sim g t_{\rm grow}$ (where g is the gravitational acceleration) which can be significantly lower than the `usual' drag terminal velocity. As for the wind tunnel simulations discussed above, several open questions remain for infalling clouds, e.g., the impact of magnetic fields on the dynamics and the survival criterion (see recent work by \citealp{Kaul2025} in this direction).

\paragraph{Other `cloud-wind' setups}
Special cases of `cloud–wind' interactions include wind tunnel setups in which the cold gas is injected through different mechanisms, i.e., not solely due to (magneto-)hydrodynamic interactions. This includes, for instance, simulations of ram pressure stripping of galaxies, where the ram pressure is counteracted by gravitational forces \citep{Gunn1972}. Numerous studies have explored such systems, with some focusing on the stripping physics itself using smaller simulation boxes \citep[e.g.,][]{Roediger2006,Choi2022}, while others employ larger domains to study the long-term evolution and multiphase structure of the stripped tail \citep[e.g.,][]{Tonnesen2010,Tonnesen2021}.
Interesting recent developments include simulations showing that the presence of a surrounding CGM around an infalling galaxy can promote additional condensation within the stripped tail \citep{Lucchini2020,Ghosh2024,Sparre2024}. Similarly, the inclusion of magnetic fields tends to suppress small-scale structure and results in smoother, less clumpy tails \citep{Ruszkowski2014}. While these ram pressure stripping simulations inherently involve multiphase gas dynamics, their primary focus is often on the global outcome—such as integrated gas loss or the impact on star formation—rather than the small-scale evolution of the multiphase interface.

Another set of wind tunnel simulations addresses the injection of relatively cold ($\lesssim 10^4\,\mathrm{K}$) stellar winds into a hot ($\gtrsim 10^6\,\mathrm{K}$), and potentially moving, ambient medium. Several studies have examined AGB winds, particularly in the case of Mira within the Local Bubble \citep[e.g.,][]{Wareing2007,Li2019}, finding good agreement with the observed tail morphology. Other works have investigated the evolution of faster, hotter winds launched by Wolf–Rayet stars \citep[e.g.,][]{Larkin2025}.

These wind–ISM interaction simulations offer yet another example of multiphase gas dynamics within galaxies. While the driving mechanism is different from classical cloud-crushing setups, many of the same physical principles—such as turbulent mixing, radiative cooling, and surface instabilities—play a role and the scalings discussed above often remain applicable.

\paragraph{Resolution requirements}
A long-standing debate in the `cloud-crushing' literature concerns what exactly constitutes `convergence' and under which conditions it is achieved. \citet{Klein1994} noted that in their (2D, adiabatic) simulations a resolution of $\sim 100$–$200$ cells per cloud radius is required for ``accurate results''. This was quantified further by \citet{Nakamura2006} (their figure~1), who showed that the cloud shape and displacement converge to within $10\%$ of their highest-resolution simulation\footnote{While one is usually constrained to comparisons against a higher-resolution numerical run, \citet{Klein2000,Klein2003} attempted to benchmark `cloud crushing' using Nova laser experiments to mitigate this limitation.} for $\gtrsim 100$ cells per cloud radius. 
However, for radiatively cooling clouds, \citet{Yirak2010} analyzed the convergence of the same global quantities and found no clear convergence (up to $1536$ cells per cloud radius), which they attributed to changes in the radiative shock structure.

In three-dimensional, radiative simulations, several studies have performed systematic resolution scans, typically spanning $8$ to $128$ cells per cloud radius \citep[e.g.][]{Cooper2009,Scannapieco2015a,Schneider2016}. Their main conclusion is that while the detailed morphology \citep[e.g., the number of fragments;][]{Cooper2009} remains unconverged, the overall mixing rate of the hot and cold phases does converge—although the evolution can be stochastic and non-monotonic with resolution.  
Based on this, studies that find cold-gas mass growth report relatively modest resolution requirements: as few as $\sim 8$ cells per cloud radius appear sufficient to capture the mass growth \citep[e.g.][]{Gronke2018, Abruzzo2022}. However, these convergence tests were typically carried out for transonic ($\mathcal{M}\sim 1.5$) flows and fiducial overdensities ($\chi\sim 100$). \citet{Gronke2022} showed that at higher Mach numbers convergence in mass growth becomes harder to achieve, although a systematic study is still lacking.
As for the turbulent mixing layers discussed in \S~\ref{sec:tmls}, convergence in the temperature distribution is much harder to achieve and likely requires the \citet{Field1965} length to be resolved \citep{Abruzzo2022}.

An important open question for larger–scale simulations is how these resolution requirements map onto the emerging picture of a critical radius, $r_{\rm crit}$, below which cold gas is rapidly mixed and above which it can grow. If $r_{\rm crit}$ indeed sets the minimum size of long–lived structures, then failing to resolve it may lead to qualitatively incorrect long term behavior. In this sense, $r_{\rm crit}$ might provide a physically motivated convergence criterion essential for accurately representing cold gas survival and growth.

\paragraph{Summary}
In summary, cloud-wind simulations have played a central role in advancing our understanding of multiphase gas dynamics. They have revealed and underlined the importance of turbulent mixing, radiative cooling, magnetic fields, and geometry in shaping the evolution band survival of cold gas in hot flows. These simulations have led to the formulation of survival criteria, uncovered mechanisms for cold gas growth and entrainment, and provided key insights that underpin modern subgrid models of multiphase gas (see \S~\ref{sec:subgrid}). Nonetheless, many open questions remain. These include the influence of realistic wind environments, the interplay between multiple physical processes, and the behavior of infalling -- as opposed to outflowing -- clouds. As simulations continue to grow in sophistication and realism (aided by computational advances), cloud-wind setups will remain an essential testbed for probing the physics of multiphase gas flows in and around galaxies.

\begin{figure}
  \centering
  \includegraphics[width=.9\linewidth]{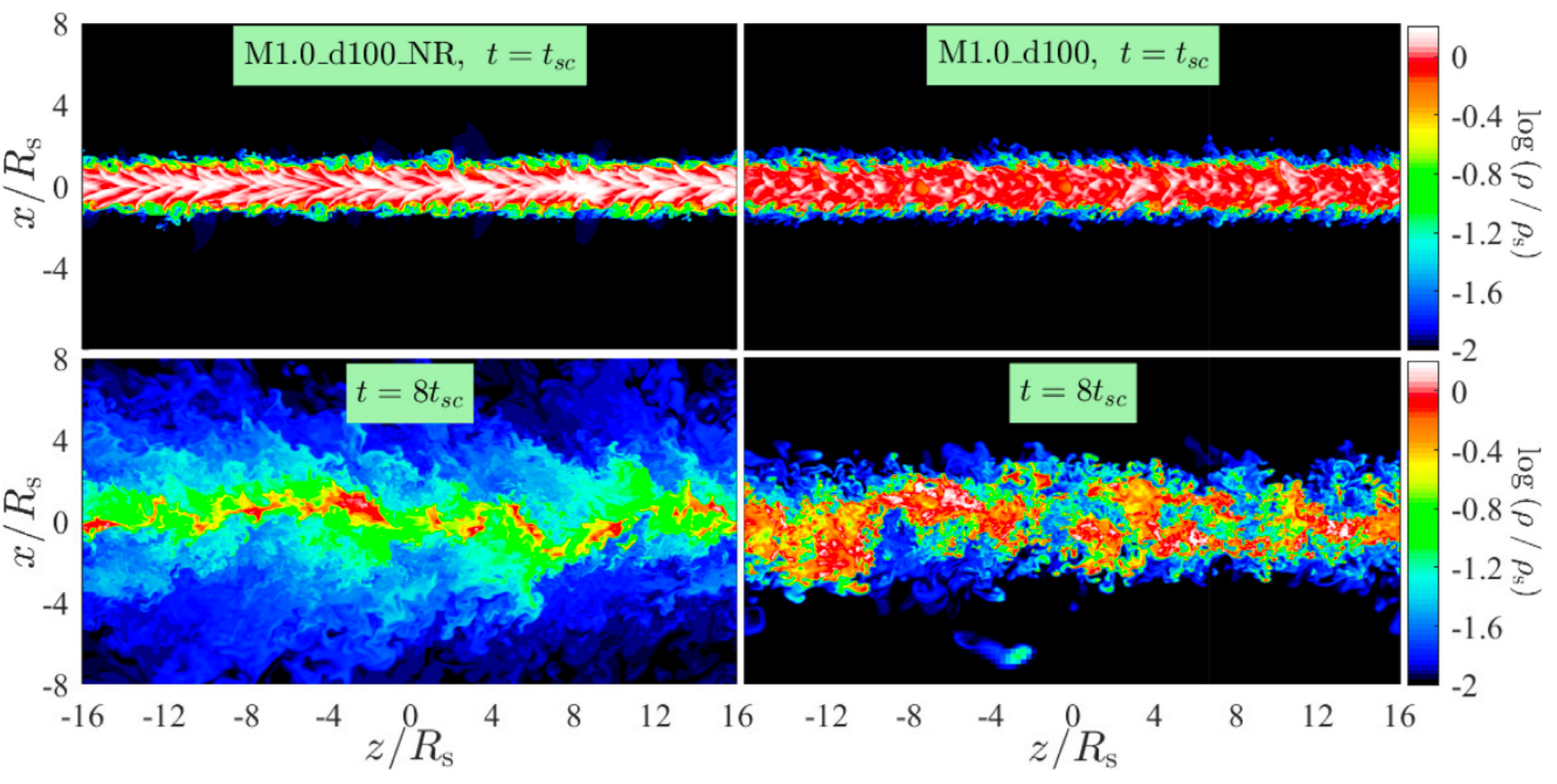}
  \caption{Stream evolution with (left) and without (right) radiative cooling. Clearly, in the simulation with cooling the dense stream stays intact wheres without cooling the mixed gas quickly dominates. Shown are density slices normalized by the initial overdensity at $1$ and $8$ sound crossing times. Adapted from \citet{mandelkerInstabilitySupersonicCold2020}.}
  \label{fig:stream}
\end{figure}

\subsection{Streams and Jets}
\label{sec:streams}

Cold gas streams penetrating hot gas halos are thought to be essential fueling mechanisms for galaxies around cosmic noon \citep{birnboimVirialShocksGalactic2003,Dekel2006}. While these streams are seen in some cosmological simulations (see \S~\ref{sec:cosmo}; \citealp{Keres2005}) their existence is heavily debated, in part because the low resolution in the halos of these large scale cosmological simulations is insufficient to capture disruption of the streams due to hydrodynamic instabilities \citep{Nelson2013}. Thus, dedicated, high-resolution, simulations of streams are often used to fill this gap. Notably, \textit{propagating} jets
and streams have a similar setup: a semi-infinite cylinder with radius $r_{\mathrm{c}}$ and density $\rho_{\rm c}$ is placed in a background medium with density $\rho_{\mathrm{b}}$ and relative velocity $v$ parallel to the cylinder. This streamwise direction is usually  periodic. The difference between what is usually considered a `jet' and a `stream' is that for the former $\rho_{\mathrm{c}} < \rho_{\mathrm{b}}$ whereas for the latter the densities are reversed. As most jet studies deal less with the multiphase dynamics of the problem, we focus in this section on streams.

The `classical' studies of adiabatic streams naturally have significant overlap with `turbulent mixing layers' (see \S~\ref{sec:tmls}) with the important difference that at high Mach numbers ($\mathcal{M}>\mathcal{M}_{\mathrm{crit}}$; cf. Eq.~\eqref{eq:Mcrit}) the streams are not stable but instead develop unstable body (or reflective) modes which can lead to similar mixing and destruction of the stream \citep{Payne1985,Hardee1988,Padnos2018}. While there are several ways a stream can be defined as `destroyed', i.e., is mixed into the surrounding hot medium, this destruction occurs approximately on a timescale of 
\begin{equation}
    t_{\rm disrupt} \sim \alpha r_{\mathrm{c}}/v
\end{equation}
with $\alpha$ being an experimentally determined fudge factor depending on $v/(c_{\rm s, hot} + c_{\rm s, cold})$ \citep{Dimotakis1991}\footnote{Since $\alpha \sim 10$ for $\chi \sim 100$, $t_{\rm disrupt}\approx t_{\rm cc}$ (cf. Eq.~\eqref{eq:tcc}) for the most commonly considered overdensity, and thus, the same `survival criterion' as Eq.~\ref{eq:tcoolmixtcc} and $r_{\rm crit}$ as discussed in \S~\ref{sec:cc} holds. However, for $\chi \sim 1$ \citet{Padnos2018} obtained $\alpha \sim 4-10$, i.e., considerably larger than $\chi^{1/2}$.}.
This non-linear disruption can be slowed down in the presence of magnetic fields \citep{Berlok2019}, thermal conduction \citep{Ledos2023}, or self-gravity \citep{Aung2019}.

When radiative cooling is included, the dynamics of streams and jets become more complex. Radiative streams can develop a multiphase structure, with cold gas cores surrounded by warmer, turbulent layers. Similar to the cloud-wind interactions discussed above, a `survival criterion' exists for which the radiative cooling is efficient enough to prevent cold gas disruption (see Figure~\ref{fig:stream}). Specifically, \citet{mandelkerInstabilitySupersonicCold2020} found this criterion to be
\begin{equation}
  \label{eq:surv_crit_streams}
  t_{\rm cool,mix} < t_{\rm disrupt},
\end{equation}
i.e., equivalent to Eq.~\eqref{eq:tcoolmixtcc} except the different form of the destruction timescale discussed above.
This survival condition can be recast in terms of a critical radius, analogous to the discussion in \S~\ref{sec:cc}, below which cold gas is mixed and destroyed faster than it can radiatively cool. In this formulation, only stream segments with $r \gtrsim r_{\rm crit}$ can persist and potentially grow. This has potentially important implications for galaxy–scale and cosmological simulations: if $r_{\rm crit}$ is not resolved, the modeled streams may artificially disrupt or, conversely, survive too easily, leading to qualitatively incorrect predictions for cold inflow, fueling, and CGM structure.
\citet{mandelkerInstabilitySupersonicCold2020} also studied the longer term evolution of the cold stream and found the inflow rate onto the stream to be $\propto t_{\rm cool}^{-1/4}$, in agreement with the turbulent mixing layer studies (see \S~\ref{sec:tmls}).

The interaction of streams and jets with their surroundings—such as the ISM, CGM or ICM is also critical and includes interesting multiphase dynamics. For instance, when a jet collides with an ambient medium (e.g., the ISM), shock fronts and contact discontinuities develop, creating a complex interface characterized by turbulent mixing and the generation of secondary instabilities \citep{Sutherland2007,Wagner2012}.
These situations are particularly relevant for the efficiency and impact of AGN feedback in the ICM/CGM and are an area of active research \citep[e.g.][]{Weinberger2023,Borodina2025}.

\subsection{Thermal instability \& Turbulent boxes}
\label{sec:turbulent}

Thermal instability is a straightforward way to create a multiphase medium: local overdensities cool slightly faster than the (thermally stable) background, increase in density, and thus cool even faster \citep[see \S\ref{sec:where_multiphase_gas_exists};][]{Field1965}. The outcome of this process is commonly referred to as \textit{precipitation} in a multiphase medium. However, the details are much more intricate and it was long though that in a stratified hot medium with a convectively stable atmosphere, condensation cannot occur because of insufficient time before large scale inflows advect the perturbations \citep{Balbus1989,Balbus1995}. Later fundamental work by \citet{Sharma2012,Mccourt2012} demonstrated, however, that if these inflows are supressed by heating, thermal instability is possible under the condition
\begin{equation}
  \frac{t_{\rm cool}}{t_{\rm ff}}\lesssim \mathcal{O}(1)
  \label{eq:tcool_tff}
\end{equation}
i.e., if the ratio of the cooling over the free-fall time is less than a constant of order unity.

The topic of thermal instability has been thoroughly discussed in the recent review by \citet{Donahue2022} (also see \citealp{Faucher-Giguere2023a}). Here, we will focus on the results of recent numerical studies. In these, the criterion mentioned above can be altered, adding various layers of complexity:
\begin{itemize}
\item  If one considers fast uplift of a gas parcel, the cooling time stays effectively constant while the free fall time dramatically increases, thus potentially fulfilling Eq.~\eqref{eq:tcool_tff} and leading to thermal instability. This was discussed in early work \citep{Shapiro1976a,Bregman1980} and more recently demonstrated in larger, halo scale simulations (e.g., \citealp{LiMODELINGACTIVE2014a}; see \S~\ref{sec:haloscale}).
  
  \item Similarly, if a halo is rotating, the effective free-fall time of a gas parcel is longer, thus, making the halo more prone to thermal instability \citep{Sobacchi2019}.

  \item The numerator of Eq.~\eqref{eq:tcool_tff} can be affected in more realistic atmospheres. As shown by \cite{Choudhury2019}, local overdensities can lower the cooling time and hence increase the probability of precipitation. In such a case, while Eq.~\eqref{eq:tcool_tff} holds locally, i.e., taking the overdensity into account in evaluating the cooling time, the global precipitation criterion (evaluated at a given radius) exhibits a lower threshold.

  \item These local overdensities can be formed naturally by turbulence. \citet{Gaspari2018} suggested using the ratio $t_{\mathrm{cool}}/t_{\mathrm{eddy}}\lesssim 1$ where $t_{\mathrm{eddy}}$ is the eddy turnover time at some (relevant) scale in the ICM (also see, e.g., \citealp{Hennebelle1999,Kritsuk2017NJPh...19f5003K,Colman2025} in the ISM context where -- while the physical conditions can be different -- an analogous criterion applies).
  Recently, \citet{Mohapatra2023} studied this criterion in conjunction with Eq.~\eqref{eq:tcool_tff} and found a new, empirical relation combining the two\footnote{Note that this precipitation criterion is different from the `survival' criterion of continued existence of multiphase medium in a turbulent medium which relates to the cold gas destruction timescale and is therefore more akin to Eq.~\eqref{eq:tcoolmixtcc} \citep{Gronke2022}.}. 

\item Magnetic fields can also alter the evolution of thermal instability. Notably, \citet{Ji2017} found that even very weak magnetic fields (as low as $\beta\sim 100$) make it easier for the gas to precipitate because of the suppression of buoyant oscillations on scales below $\sim v_{\rm A}t_{\rm cool}$ (where $v_{\mathrm{A}}$ is the Alfv\'en velocity). Furthermore, they showed that the longer-term, nonlinear evolution exhibits a very different multiphase gas morphology with the cold gas being more filamentary \citep[see also][]{Sharma2010}.

  \item Another important non-thermal effect on precipitation are cosmic rays, which is still very much an active field of research\footnote{Numerical advances including the `two-moment method' \citep{Jiang2018,Chan2019MNRAS.488.3716C,Thomas2019a} have recently allowed more realistic cosmic ray transport to be included in multiphase simulations.}.
  The linear stability criterion has been studied by \citet{Kempski2020} who formulate thermal instability criteria in the presence of cosmic ray heating. The non-linear evolution has been the subject of several recent studies \citep[e.g.][]{Butsky2020,Tsung2023MNRAS.526.3301T,Weber2025} that point out the change in morphology and dynamics of the multiphase gas that forms. One obvious point, for instance, is that the non-thermal pressure support of cosmic rays can lead to significantly lower overdensities of the cold material.
\end{itemize}
    While the theoretical side of thermal instability stands on solid ground and is supported by a range of numerical experiments (some of which are mentioned above), the application of these models to larger scale, cosmological simulations is less certain  (see also \S~\ref{sec:cosmo}). For instance, studies by \citet{Esmerian2020} and \citet{Nelson2020}, using the \texttt{FIRE} and \texttt{Illustris-TNG50} simulations, respectively, show that the ratio $t_{\mathrm{cool}}/t_{\mathrm{ff}}$ of a gas parcel in a galaxy halo does not determine its thermal evolution well in a cosmological context (which could be due to the hot medium inflowing as a cooling flow in the \texttt{FIRE} simulations; \citealp{Hafen2022,Sultan2025}). Ultimately, comparison with observations will be the ultimate benchmark \citep[see][for a comprehensive overview of the current observations]{Donahue2022}. \\

\begin{figure}
  \centering
  \includegraphics[width=0.8\linewidth]{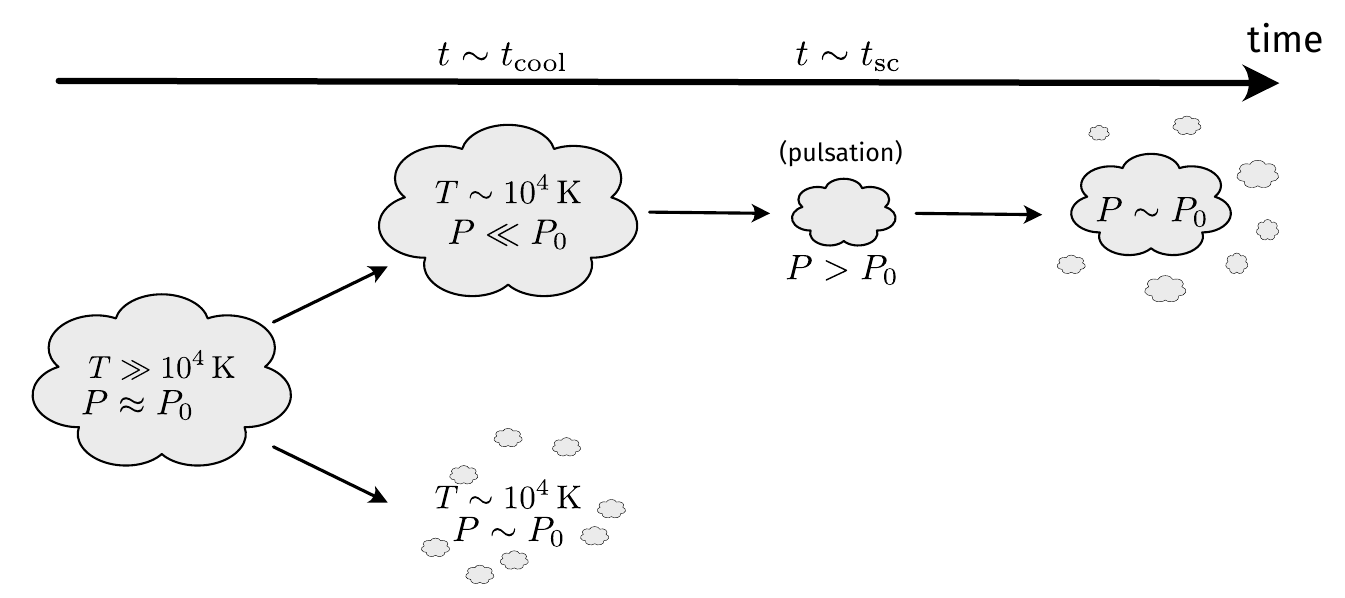}
  \caption{Illustration two possibilities on how a large ($r_{\rm cl}\gg l_{\rm shatter}$) cloud can fragment through cooling. In the upper pathway the cooling cloud is wildly out of pressure balance which leads to a violent pulsation process during which the outer layers of the cloud are being shed. In contrast, the lower pathway illustrated the option of a continuous fragmentation process to smaller and smaller clouds of size $\sim c_{\rm s}t_{\rm cool}$ which keep in pressure balance. Whereas the latter was imagined by \citet{McCourt2016} naming this process `shattering', the former seems to occur in more recent simulations \citep[][albeit with different initial conditions, see main text for details]{Gronke2022,Yao2024} and was dubbed `splattering' by \citet{Waters2019b}.}
  \label{fig:shattering_sketch}
\end{figure}

Beyond the question whether and where thermal instability will occur, a significant focus of simulations of multiphase media has been dedicated to the study of the longer term, non-linear evolution of precipitating gas and the resulting gas morphology. Early work by \citet{Hennebelle1999} and \citet{Burkert2000} pointed out the possibility of cooling aided fragmentation, and thus, the emergence of small cold clouds in a hot medium. Later, \citet{McCourt2016} showed in influential work\footnote{Around the same time, \citet{Cornuault2018} described a similar fragmentation process due to cooling in the context of streams. Notable differences to \citet{McCourt2016} include that they evaluate the cooling length $t_{\mathrm{cool}}c_{\mathrm{s}}$ not at the minimum but at the post-shock conditions, and they do not test their theory using hydrodynamical simulations.} that while gas clouds of size $r < l_{\mathrm{shatter}}$ where
\begin{equation}
  \label{eq:lshatter}
  l_{\mathrm{shatter}}\equiv \mathrm{min}(c_{\mathrm{s}}t_{\mathrm{cool}})
\end{equation}
are in sonic contact with their surroundings, and thus cool isobarically, larger clouds ($r \gg l_{\mathrm{shatter}}$) do not cool isochorically, as was commonly assumed. Rather, they are highly underpressurized between $t\sim t_{\mathrm{cool}}$ and the sound crossing time $t\sim r/c_{\mathrm{s}}$. Instead, \citet{McCourt2016} suggested that large clouds fragment while cooling into clumps eventually of size $l_{\mathrm{shatter}}$, which stay in sonic contact. They demonstrated the different resulting cold gas morphology using large, two-dimensional simulations\footnote{Note that the simulations in \citet{McCourt2016} (also) show a range of clump sizes which might be due to `imperfect fragmentation' or rapid coagulation.}, and discuss their findings in the context of observations where evidence of small-scale cold gas fragments has been found.

Since then, there has been some debate in the literature regarding the nature of cooling driven fragmentation. \citet{Waters2019b} give a counterpoint to the `shattering' picture (also summarized in \citealp{Waters2023a}) above. In this study, they focus on the linear theory of thermal instability and show using 1D simulations that large clouds can oscillate when cooling. The authors furthermore speculate that the fragmentation seen in \citet{McCourt2016} is likely merely numerical, as they have seen similar fragmentation artifacts in their 1D simulations. \citet{Waters2019b} extrapolate the oscillations they observe and conjecture that a `rebounce' is followed by fragmentation (see Figure~\ref{fig:shattering_sketch} for a visual representation of the two pathways). Indeed, such a process is seen in 3D simulations \citep{Gronke2020,Yao2024}, which also extend the fragmentation criterion $r \gg l_{\mathrm{shatter}}$ to contain an (final) overdensity dependence.
These findings point towards a dynamical picture with the cold gas clumps formed towards the end of a violent pulsation, i.e., supporting the `splattering' picture of \citet{Waters2019b}. There are, however, many outstanding questions, including:
\begin{itemize}
\item the importance of the dimensionality of the simulations. One should not forget that some of the (especially earlier) studies mentioned above were carried out in 1D, while some (cooling aided) fragmentation effects such as Kelvin-Helmholtz aided fragmentation of the surface \citep[e.g.][]{Liang2018} or fragmentation in a turbulent medium \citep{Saury2014,mohapatraTurbulenceIntraclusterMedium2019,Gronke2022} only occur in multi-D simulations\footnote{\citet{Farber2023} point out, e.g., that in fact rotation of the clumps leads to continued fragmentation after the expansion phase (an effect they dub `splintering').}. Indeed, \citet{McCourt2016} state explicitly that `shattering' cannot occur in 1D; however, note that \citet{Das2020} and \citet{Waters2019b} support and oppose the `shattering' model, respectively, using 1D simulations.
\item the effect of other important physics such as thermal conduction on the cold clump sizes. This was discussed in the literature as the Field length might set the characteristic length scale after all. However, mainly due to the uncertain effect of (tangled) magnetic fields conclusions differ widely \citep{Burkert2000,McCourt2016,Waters2019b,jenningsThermalInstabilityMultiphase2020}.
\item the role of droplet coagulation during the cooling and in the longer term process, and whether this is related to the fragmentation criterion discussed above. While potential coagulation of cold clumps in multiphase media is a long-known phenomenon \citep{ZelDovich1969,Elphick1991,Elphick1992}, only more recently have studies focused on cooling driven coagulation of clumps, i.e., the effect of being entrained in the accreting hot material \citep{Waters2019,Koyama2004,Gronke2023}. %
While this effect is found to be rather slow, it depends significantly on the distance between the droplets, and thus, might still be critical in the early stages of fragmentation.
\item the impact of the initial conditions. It is important to note that while \citet{McCourt2016} used a truncated Gaussian random field, more recent studies used simplistic (e.g., top-hat or solenoidal) initial conditions, which might lead to differences in the evolution and outcome.
\end{itemize}
As in many active fields of research, the topic of `shattering' is not yet fully addressed, and the most recent literature is rather inconclusive -- a fact that is not aided by the different definitions of terminology found in the literature. Hopefully, new simulations addressing some of the points raised above will shed light on the picture and in particular answer whether a Jeans-like continuous fragmentation can occur and whether a `characteristic scale' of multiphase gas exists, and to what extent cloud fragmentation may instead be driven by turbulent motions alone, without invoking cooling driven `shattering' mechanisms.

\subsection{Supernova- and Wind-driven Bubbles in the ISM}
\label{sec:superbubbles}

Using numerical simulations to better understand the phase structure and evolution of the ISM has a long history. Indeed, some of the earliest cloud-crushing simulations were designed to replicate the effects of a supernova shock sweeping over a small interstellar cloud (see \S~\ref{sec:cc}), such that the radiative cooling timescale was sufficiently long to be ignored \citep{Nittmann1982, Klein1994}. It was recognized early on that stellar winds and supernova explosions could provide a continuous source for the hot phase of the ISM \citep[e.g.][]{Cox1974}. The question of how the hot ($T > 10^6$~K) bubbles generated by these sources interact with the existing cooler ambient ISM phases is the focus of simulations in this section.

In the case of a single supernova explosion with energy $E_\mathrm{SN}$ and ejecta mass $M_\mathrm{ej}$ in an ambient medium with number density $n_0$, a one-dimensional solution for the resulting time-dependent evolution can be found analytically \citep{Ostriker88}. This analytic solution can be used to estimate the size, post-shock temperature, post-shock velocity, and total hot gas mass at various times during the bubble’s evolution \citep[e.g.][]{Draine2011}. A common practice is to evaluate these quantities at the time when the evolution transitions from primarily energy conserving (the Sedov-Taylor phase) to a momentum conserving phase. This occurs when the material swept up at the edge of the bubble begins to radiate efficiently — the ``shell formation time”:
\begin{equation}
    t_\mathrm{sf} \approx 4.4\times10^4\,\mathrm{yr}\,E_{51}^{0.22}n_0^{-0.55},
\end{equation}
where $E_{51}$ is the energy of the supernova in units of $10^{51}$ erg, and $n_0$ is the number density of the ambient medium in units of $1 n_\mathrm{H}\, \mathrm{cm}^{-3}$ \citep{Draine2011, Kim2015}.\footnote{Note that the exact value of the normalization in this function will depend on the specifics of the cooling function.}

Three dimensional simulations of explosions in a constant density ambient medium agree well with the analytic estimates at $t_\mathrm{sf}$ \citep{Martizzi2015, Kim2015}. Given a constant explosion energy, the size of the resulting supernova bubble, total mass of hot gas, and total radial momentum scale inversely with the ambient density, although weakly -- fits from hydrodynamic simulations including radiative cooling at solar metallicity in \cite{Kim2015} are $r_\mathrm{sf} = 22.1\,\mathrm{pc}\,n_0^{-0.43}$, $M_\mathrm{sf}~=~1550\,\mathrm{M}_\odot\,n_0^{-0.29}$, and $p_\mathrm{sf}=  2\times10^5\,\mathrm{M}_\odot\,\mathrm{km}\,\mathrm{s}^{-1} n_0^{-0.15}$ for the bubble radius, hot gas mass, and momentum respectively, which are within 10\% of the analytic estimates (small differences are expected given the simplifications made to the functional form of the cooling function in the analytic case). Importantly, \cite{Yadav2017} showed that although it does not affect the size of the bubble or speed of expansion, the inclusion of thermal conduction can significantly change the properties of the interior due to the evaporative flow that develops from the shell to the interior of the bubble, decreasing the overall interior temperature but increasing the total mass of hot gas \citep[see also][]{El-Badry2019}.

At sufficiently high pressure the ISM is known to be inhomogeneous, with a stable two-phase equilibrium consisting of a dense, cold ($T\sim 10^2$ K) neutral medium (CNM) surrounded by more diffuse, warm ($T\sim10^4$ K) gas \citep{Field1969, McKee1977}. Large enough clouds in the CNM can self-shield, allowing for even colder molecular clouds to develop, which are themselves highly inhomogenous as a result of large density contrasts inherent in supersonic turbulence. This complication motivated a series of three-dimensional computational studies of supernova interactions with a multiphase ISM \citep{Iffrig2015, Martizzi2015, Kim2015, Walch2015, Li2015b}. These models demonstrated that for the same \textit{average} ambient density, a multiphase ISM background results in supernova bubbles with comparable total hot gas and slightly lower terminal momentum as compared to explosions in a single-phase medium, though the time when significant radiative losses commence can be significantly earlier \citep{Martizzi2015, Kim2015}. This is because the embedded cold clouds are initially shielded as the supernova shock propagates through low density channels, but then begin to mix with the post-shock gas, generating high-density, efficiently cooling warm gas in the bubble interior \citep{Walch2015}. In addition, the supernova shock propagates more quickly through the low density channels in a multiphase ISM, leading to larger bubble sizes overall. 

Naturally, the morphology of the supernova bubble is also very different in the clumpy ISM case, and depends on the size scale of the cold clouds relative to the overall size of the bubble \citep{Martizzi2015}. In addition, substantial scatter in the bubble properties can be introduced depending on the exact location of the supernova relative to the initial background density perturbations, especially if the initial conditions account for the spatial correlations between density perturbations at different scales, as is expected in a realistic ISM structure with the velocity statistics of the three dimensional turbulence \citep{Ohlin2019}. The location of explosions (at density peaks versus in lower density environments) can also significantly affect the overall ISM phase structure; explosions in high density locations suffer considerably more radiative losses but generate a more turbulent cold ISM, while explosions in low density environments generate a significant and long-lived hot phase \citep{Gatto2015}. For the case where sufficiently many supernovae explode within a multiphase volume, \cite{Li2015b} showed that the overall phase structure of the ISM can transition from a warm-dominated to a hot dominated state, with the majority of the volume containing $T > 10^6$ K gas. They found that this transition happens when the fraction of the volume filled with hot gas exceeds 60\%. At this point, supernovae do not evolve as individual bubbles any more, but begin to overlap significantly, leading to thermal runaway and a wind-generating state. However, this transition is quite sensitive to the photoelectric heating (PEH) rate, and if the the PEH rate is proportional to the rate of supernova explosions, they find that the ambient medium is always stable for average number densities $\bar{n} \gtrsim 3\,\mathrm{cm}^{-3}$.

A defining characteristic of the single supernova case is that the majority of the hot gas and momentum are generated while the supernova remnant is still in the energy-conserving stage. A related class of experiments explores the effect of multiple supernovae exploding at the same location in sequence, as would be expected in e.g. a massive star cluster. These simulations of ``superbubbles'' share many similarities with the single supernova case, but are typically parametrized by an energy injection rate or cluster mass (\citealp{MacLow1988}; the two can be related by a given stellar population model). For the case of multiple supernovae exploding in a periodic box larger than the radius of the generated bubble, a primary difference from the single supernova case is that the total hot gas generated per supernova explosion is lower, as the bubble is effectively in the momentum-driven limit for most of its evolution \citep{Martizzi2015, Kim2017, Fielding2018}. However, the total radiative energy losses can be significantly reduced if the bubble grows larger than the ISM scale height and is able to break out, as will often be the case for large clusters in galactic disks \citep{Fielding2018}.

Using 1D simulations and a homogeneous ISM, \cite{Gentry2017} argued that the total momentum budget per supernova for clustered supernova explosions can be as much as an order of magnitude higher than for the single supernova case, a result that contradicts the three-dimensional simulations described above, which found comparable terminal momentum per explosion to the single supernova case. In followup work using 3D simulations and a meshless-finite-mass hydrodynamic scheme, \cite{Gentry2019} found a final momentum more similar to the other 3D results, but they noted that the overall momentum evolution depended on how well mixing at the bubble interface was resolved; increasing resolution of instabilities yielded progressively higher total momentum. This may reflect a difference in numerical algorithm however, as other studies using finite volume methods did not find such a dependence, and instead found that their results were converged at resolutions of several parsecs \citep[see Appendix,][]{Kim2017}.

Closely related to superbubbles are stellar wind-blown bubbles, which evolve similarly in the case when supernovae are closely spaced in time, but with lower total energies \citep{Weaver1977}. A primary conclusion from 3D simulations of these winds is that the turbulent mixing at the fractal surface of the bubble allows for much more efficient radiative losses at the bubble-ISM interface, fundamentally altering the solution from an energy driven to a momentum driven case \citep{Lancaster2021b, Lancaster2024}. The theory describing the cooling rate at the superbubble interface is closely related to the that of turbulent radiative mixing layers (cf. \S~\ref{sec:tmls}).

\subsection{ISM Patch -- Models of Stratified Disks}
\label{sec:ISMpatch}

Most of the simulations discussed in the previous section were focused on the evolution of supernova- and stellar wind-driven bubbles within the ISM, and therefore were carried out in small volumes ($\lesssim 1\,\mathrm{kpc}^3$) at high spatial resolution ($\lesssim 1\,\mathrm{pc}$), and neglected the effect of a background gravitational potential. In order to capture the evolution of stellar feedback on larger scales and the emergence of galactic winds, simulations that also incorporate the effects of gravity are required. We collectively refer to this class as ``stratified disk'' or ``ISM Patch" simulations, and will distinguish in the next two subsections between those that include a relatively small subset of physics and a more idealized setup (\S \ref{subsec:simpledisks}), and those that include a larger range of physical processes (particularly gas self-gravity) and attempt to capture the evolution of the ISM, star formation, and supernova feedback self-consistently (\S \ref{subsec:complexdisks}). Also known as ``tall box" simulations, most of the following setups use boxes with shorter $x$ and $y$ dimensions (500 pc or 1kpc are typical) and periodic or shearing $x$- and $y$-boundaries, and longer $z$ dimensions (ranging from 1 to 10 kpc) with outflow boundaries in the $z$-direction (see Figure \ref{fig:simulations}).

In the following results, we will make use of the fact that given a star formation rate, many features of outflows can be framed in the context of ``loading factors''. For example, the mass outflow rate in a given phase can be parametrized relative to the star formation rate as
\begin{equation}
    \eta_\mathrm{m} = \frac{\dot{M}_\mathrm{outflow}}{\mathrm{SFR}}.
\end{equation}
This is known as the ``mass loading factor'', and is a useful way to normalize outflow rates across a range of different galaxy parameters.
Similarly, one can calculate an ``energy loading'':
\begin{equation}
    \eta_\mathrm{E} = \frac{\dot{E}_\mathrm{outflow}}{\dot{E}_\mathrm{SN}},
\end{equation}
where $\dot{E}_\mathrm{SN}$ relative to the star formation rate will depend on the exact nature of the chosen IMF and supernova explosion energy (typically assumed to be $10^{51}\,\mathrm{erg}$).

These loading factors as well as other statistics like volume-filling fractions are frequently measured for different phases in the simulations (for example, see the phase diagram in Figure~\ref{fig:coolingcurve}). We will define the different phases using the following approximate temperature cuts (though note that these can vary slightly from one simulation to another): hot: $T > 5\times10^5$ K, intermediate: $5\times 10^5\,\mathrm{K} < T < 2\times 10^4\,\mathrm{K}$, warm: $2\times 10^4\,\mathrm{K} < T < 5000\,\mathrm{K}$, unstable: $5000\,\mathrm{K} < T < 300\,\mathrm{K}$, and cold: $T< 300\,\mathrm{K}$. The cold, warm, and hot phases are the traditionally stable ranges in the three-phase ISM model of \cite{McKee1977}, while the intermediate and unstable phases are generally transient and must constantly be replenished.

\subsubsection{Simplified setups}
\label{subsec:simpledisks}

Following pioneering early work by \cite{Korpi1999}, \cite{deAvillez2004} carried out some of the first high-resolution 3D (AMR) simulations of a vertically-stratified ISM, incorporating the effects of radiative heating and cooling, as well as magnetic fields \citep{deAvillez2005}. A major focus of their work involved investigating the discrepancy between the predicted ISM hot phase filling fraction, $f_\mathrm{v, hot}$, of ~75\% \citep{McKee1977}, and observed constraints for the Milky Way and nearby galaxies, which indicated much lower values ($f_\mathrm{v, hot} < 0.5$). Using a vertically-stratified setup that extended significantly above and below the disk midplane, they demonstrated that the ability of supernova bubbles to vent out of the ISM and into the halo provides a ``release valve” that constrains the typical $f_\mathrm{v, hot}$ to less than 50\%, and found only of order 20\% for solar-neighborhood-type conditions and supernova rates (a result that is robust to the presence of a galactic magnetic field). Similar work by \cite{Joung2006} found higher $f_\mathrm{v, hot}$ of 0.4-0.5 for Milky-Way-like conditions, and they pointed out that the value is quite sensitive to the assumed heating rate, since a higher photoelectric heating (PEH) rate shifts much of the ISM from the cold neutral medium (CNM) into the warm phase, effectively increasing the volume-filling ambient density and subsequently decreasing $f_\mathrm{v, hot}$. They also showed that given a realistic PEH rate, a significant fraction of gas can exist in the thermally unstable regime ($200\,\mathrm{K} < T < 10^4\,\mathrm{K}$), consistent with observations.

In more recent work using a similar vertically-stratified disk setup, \cite{Martizzi2016} demonstrated that the overall efficiency of supernovae in generating hot outflows is strongly dependent on the vertical location of the explosions. In models that seeded supernovae at $z$ locations strictly proportional to the gas density (and thus preferentially near the midplane), the resulting outflows were significantly less energetic than models that randomly exploded supernova within a given scale-height, $h_\mathrm{SN}$, (as might reflect a large number of supernovae occurring from runaways). This random seeding resulted in a much higher fraction of explosions occurring in lower density environments and proved much more efficient at generating hot winds that can escape the ISM. In a similar study, \cite{Li2017} found that the total energy and metal mass incorporated into the escaping hot phase is enhanced at higher $h_\mathrm{SN}$. These models where the supernova scale-height is treated independently from the gas disk scale height can thus result in the somewhat counterintuitive effect that at fixed gas surface density, an increased gravitational potential increases the efficiency of supernovae at generating winds, by decreasing the scale height of the warm ISM. These results were consistent with earlier work by \cite{Creasey2015}, who argued that the correlation between increased surface density and hotter, more metal-enriched outflows could explain the flattening of the mass-metallicity relation at higher galaxy masses.

Tying this effect back to the analytic estimates for superbubble evolution, \cite{Fielding2018} parametrized the potential for hot wind generation using star cluster mass, showing that when clusters are massive enough (such that the superbubble can grow to larger to the ISM scale-height in less than $\sim30$ Myr), bubbles can break out of the ISM and vent freely into the surrounding low-density halo. In particular, for a disk with scale height $h$ and gas surface density $\Sigma_\mathrm{g}$, this will happen if the star cluster formation efficiency, $\epsilon_* = M_\mathrm{cl}/(\pi h^2 \Sigma_\mathrm{g})$, exceeds a critical value:
\begin{equation}
\epsilon_{*,\mathrm{crit}} = 0.015 \left(\frac{f_\mathrm{V}}{0.25}\right)\left(\frac{\delta v}{10\,\mathrm{km}\,\mathrm{s}^{-1}}\right)^2\left(\frac{h}{100\,\mathrm{pc}}\right)^{-1}\left(\frac{p_\mathrm{SN}}{10^5\,\mathrm{M}_\odot\,\mathrm{km}\,\mathrm{s}^{-1}}\right)^{-1},
\end{equation}
where $f_\mathrm{V}$ is the ratio of the median ISM number density to the mean, $\delta v$ is the typical turbulent velocity dispersion of the ambient medium, and $p_\mathrm{SN}$ is the momentum injected per supernova. Below this value, bubbles tend to ``stall out", as a result of the turbulence in the interstellar medium, while above it, they are able to break out. This corresponds to a star formation rate surface density threshold of $\sim 0.03\,\mathrm{M}_\odot\,\mathrm{yr}^{-1}\,\mathrm{kpc}^{-2}$, or a molecular gas surface density threshold of $\Sigma_{\rm H2} \sim 60\,\mathrm{M}_\odot\,\mathrm{pc}^{-2}$, assuming an $\mathrm{H}_2$ depletion time of 2 Gyr. In this case, much of the energy ($\sim$10-50\%) associated with explosions post-breakout is able to contribute to the generation of a hot, low density, fast outflow, significantly altering the evolution from the radiatively efficient, momentum-driven limit found at late times in periodic box simulations \citep{Kim2017}.

\cite{LiM2020} summarized the findings of many of these stratified-disk models across a range of star formation rate surface densities, $\Sigma_\mathrm{SFR}\sim [10^{-4}, 1]\,\mathrm{M}_\odot\,\mathrm{kpc}^{-2}\,\mathrm{yr}^{-1}$ and gas surface densities, $\Sigma_\mathrm{gas}\sim [1,300]\,\mathrm{M}_\odot\,\mathrm{kpc}^{-2}$ \citep{Creasey2015, Li2017, Martizzi2016, Fielding2018, Armillotta2019}. Among the main conclusions are:
\begin{itemize}
\item The hot phase ($T\sim 10^6$ K) has low mass loading, $\eta_\mathrm{m} \sim 0.1-0.2$, but dominates the energy flux of outflows.
\item The specific energy of the hot phase, $e_{s, h}$, is only weakly correlated with the star formation rate surface density as $e_{s,h} \propto \Sigma_\mathrm{SFR}^{0.2}$, and has a relatively narrow range.
\item The specific energy of the hot phase is typically much higher than the specific energy of the warm phase ($T\sim10^4$ K), by factors of 10 - 1000. This implies that for a given gravitational potential, the high specific energy hot outflows can travel much farther than the low specific energy warm phase, so hot outflows likely play an important role in regulating halo properties and thus galaxy star formation rates over long timescales.
\end{itemize}

More recent simulations have corroborated many of these results. Using a modified version of the \cite{Fielding2018} setup and including MHD, \cite{Tan2024a} demonstrated that clustered supernovae in a turbulent, multiphase ISM lead to asymmetric and time-variable breakout, and that a population of $T\sim 10^4$ K clouds is seeded in hot outflows as a result of the resulting warm ISM fragmentation. In some of the largest fixed-resolution tall-box simulations to-date, \cite{Vijayan2024} demonstrate that the hot phase of outflows should be the most metal-rich \citep[see also][]{Creasey2015, Li2017}, and that the exchange of gas between phases within outflows should result in decreasing hot phase metallicity with height, and increasing warm gas metallicity. In particular, they note that effects that depend on the exchange of mass between phases (such as the metal loading), are only converged at high (2 - 4pc) resolution.

\subsubsection{Simulations including self-gravity}
\label{subsec:complexdisks}

In the simulations described in the previous subsection, a key goal was tunability — these experiments were typically designed to vary a single feature (gravitational potential, presence or absence of magnetic fields, etc.) in a simplified setup while holding other features fixed, and thus determine how that variable affected the resulting multiphase structure of the ISM and/or outflows. An important variable in these models was the occurrence rate and location of supernova explosions, which were set ad hoc. In this subsection we discuss a complimentary group of simulations that we refer to as ``self-consistent'', which seek to capture the primary physical processes of the ISM (at a minimum, typically star formation and supernova feedback), and describe the resulting ISM and outflow structure, often with the goal of developing analytic models of these relationships \citep[e.g.][]{Ostriker2011, Faucher-Giguere2013, Ostriker2022}. We refer to these simulations as self-consistent because their goal is to capture star formation (and resulting supernovae) in a manner that depends directly on the changing features of the local ISM and realistic stellar populations. Nevertheless, we note that even in this case, assumptions and algorithmic decisions must be made that can impact the results, such as whether to include Type Ia supernovae, or whether to seed star particles as clusters (which may artificially increase clustering of supernova events) or individual massive stars (which may artificially decrease clustering in particle-based methods, due to finite gravitational softening lengths).

Many of these simulations also incorporate increasingly complex additional physics, including magnetic fields, chemical evolution, dust, stellar winds and radiative feedback, cosmic rays, and more. While these models are thus in many ways more realistic than those discussed in the previous section, they also involve more complex interrelationships between various physical processes, making it more challenging to derive clear relationships between a particular variable and the properties of multiphase gas. These simulations were carried out with many different goals, and it is beyond the scope of this review to discuss all of the advances they have made. We will therefore attempt to emphasize primarily those results that focus on the multiphase nature of the ISM and outflows.

\begin{figure}
\centering
\includegraphics[width=0.9\linewidth]{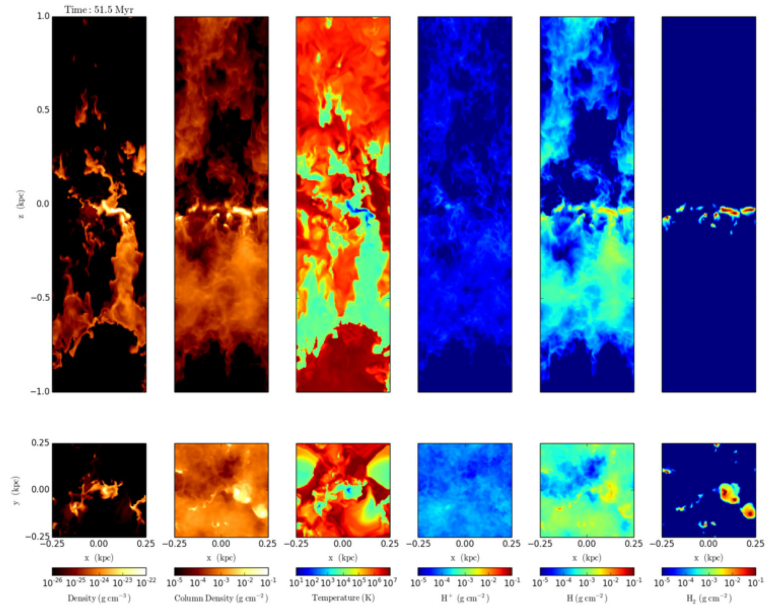}
\includegraphics[width=0.95\linewidth]{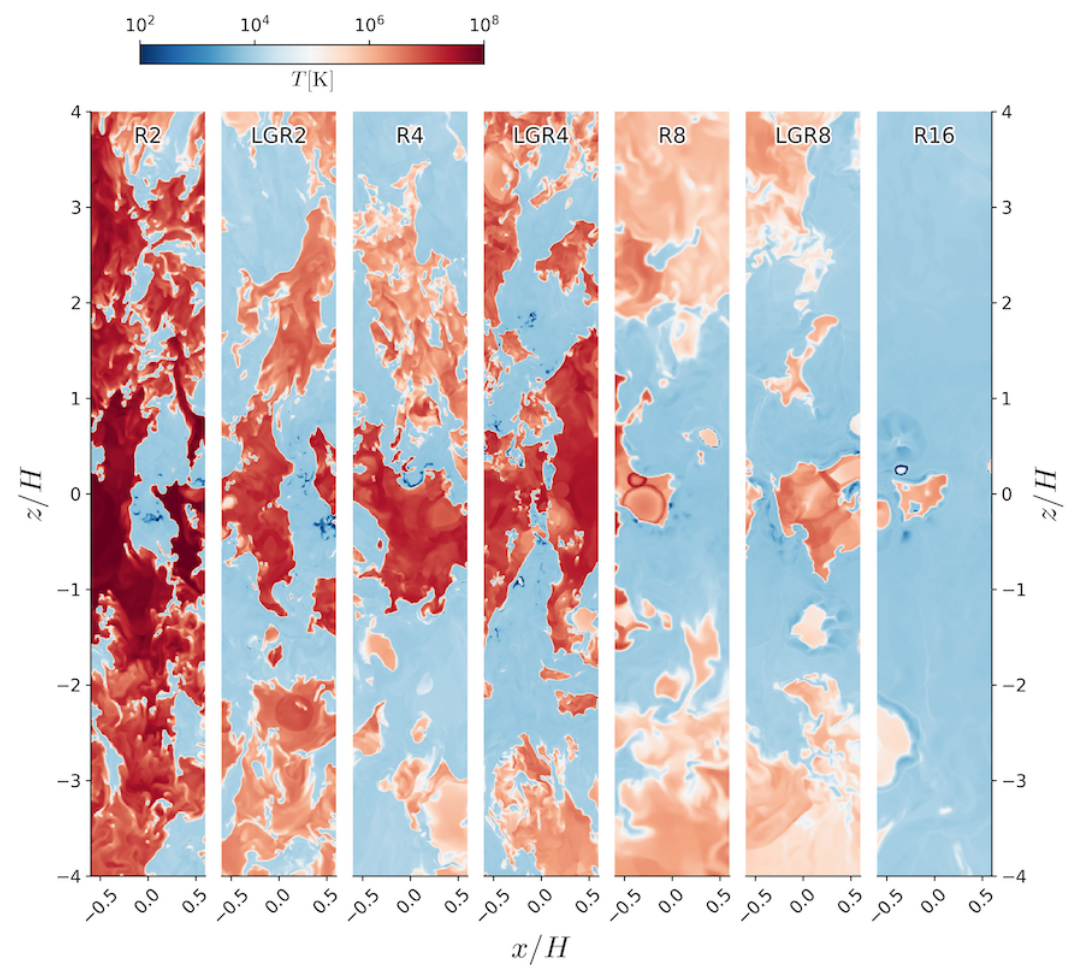}
\caption{Snapshots from the SILCC \citep{Walch2015} and TIGRESS \citep{Kim2020} simulation suites (top and bottom, respectively), showing the generation of multiphase outflows from the ISM driven by supernova feedback. The top panel shows snapshots of the same simulation in different variables, while the bottom panel shows temperature slices through seven simulations with different gas surface densities and gravitational potentials.}\label{fig:ISM_patch}
\end{figure}

As in the simulations discussed in Section \ref{subsec:simpledisks}, the scale of these simulations is typically set by a competition between the need to resolve processes occurring on the scale of star clusters (a few pc) and capture evolution on scales relevant to integrated properties of galaxies (kpc). Thus, a common approach is to use a ``tall box" setup -- a $\sim0.5\,\mathrm{kpc}^2$ patch of a galaxy that extends several kpc above and below the disk (see Figure \ref{fig:ISM_patch}), and employ sub grid models for star formation (so called ``sink particles” which accrete gas and turn into star clusters) and supernova momentum injection in cases where the Sedov-Taylor phase cannot be fully resolved (see e.g., \citealp{KimCG2013, Hennebelle2014, Walch2015} for descriptions of feedback prescriptions, and \citealp{Hennebelle2014, Gatto2017, Kim2017} for examples of sink particle creation).

Here we highlight results mainly from mature simulation suites that have largely attempted to model Milky Way-like galaxies at solar metallicity, with the earliest work in particular attempting to replicate conditions in the solar neighborhood. More recent models have expanded this parameter space somewhat, though a more prominent focus has been on expanding the range of physical processes that are modeled. Some of the key results of these complex, multi-faceted studies include:
\begin{itemize}
\item Self gravity is key to accurately capturing the formation of cold gas and molecular hydrogen, and including self gravity increases the hot gas volume filling fraction in simulations \citep{Walch2015}.
\item Magnetic fields decrease the efficiency of molecular gas formation on short time scales ($< 30$ Myr) \citep{Walch2015} and therefore the star formation rate \citep{Hennebelle2014}, but do not have a significant dynamical impact \citep{Girichidis2016, Kim2017b}.
\item A significant fraction of supernovae must explode in low density environments to develop a multiphase ISM structure that is consistent with observations, including velocity dispersions of various components \citep{Walch2015, Girichidis2016}. This can be achieved via spatial and temporal clustering of explosions in clusters, runaway OB stars, Sne Type 1a, or significant pre-supernova feedback \citep[e.g.][]{Peters2016}.
\item Thermally-driven outflows typically do not contain cold molecular gas, and rather by mass are mostly composed of warm atomic hydrogen at $T\sim 10^4$ K \citep{Girichidis2016, Kim2018}.
\item Stellar winds (and other pre-supernova feedback, including radiation) lower the mass of forming clusters \citep{Gatto2017, Rathjen2021}, which can in turn decrease the efficiency of supernova in driving outflows. On the other hand, pre-supernova radiative feedback can significantly lower the ambient density in which supernovae explode, potentially increasing the impact of explosions outside of clusters \citep{Peters2016}. In the densest regions, radiation is more important than stellar winds for disrupting the ambient ISM \citep{Haid2019}.
\item At solar neighborhood conditions, most warm gas doesn’t achieve high enough velocities to escape as a wind and rather forms a fountain \citep{Girichidis2016, Kim2017b}, with outflow rates comparable to the star formation rate at heights of $\sim1$ kpc, but significantly lower ($\eta_m\approx0.03$) at heights of 4 kpc or so \citep{Kim2018}.
\item Because warm gas can have a large range of outflow velocities, some of which are consistent with a fountain flow rather than a wind, it is better characterized by a velocity distribution than a single mass-loading factor \citep{Kim2018, Rathjen2023}.
\item At solar neighborhood conditions, thermal supernova-driven outflows have hot gas outflow mass loading factors of order $\eta_\mathrm{m, hot} \sim 0.1$, and energy loading factors of order 2\% \citep{Kim2018}. These hot gas mass loadings are fairly flat ($\eta_\mathrm{m, hot} \sim 0.1-0.2$) across a broad range of $\Sigma_\mathrm{SFR}$ ($10^{-4} - 1\,\mathrm{M}_\odot\,\mathrm{kpc}^{-2}\,\mathrm{yr}^{-1}$) \citep{Kim2020}, which is consistent with predictions from the more idealized simulations discussed in \S \ref{sec:superbubbles} and \S\ref{subsec:simpledisks}.
\item Hot gas energy loading is a weakly increasing function of $\Sigma_\mathrm{SFR}$, likely due to increased spatial and temporal clustering of Sne with increasing SFR. Conversely, warm gas energy loading is a weakly decreasing function of $\Sigma_\mathrm{SFR}$, while warm gas mass loading is a steeply decreasing function of $\Sigma_\mathrm{SFR}$. \cite{Kim2020} find $\eta_\mathrm{m, hot} \propto \Sigma_\mathrm{SFR}^{0.14}$ \citep[also see][]{Li2020}, $\eta_\mathrm{E, warm} \propto \Sigma_\mathrm{SFR}^{-0.12}$, and $\eta_\mathrm{m, warm} \propto \Sigma_\mathrm{SFR}^{-0.44}$, however most of this gas is not high enough velocity to escape (see above).
\end{itemize}

Although the exact quantitative results in many of the points above differ slightly from one simulation to another, by-and-large the picture that has emerged is consistent. However, one area where simulations still exhibit some significant differences regards the volume-filling fraction of hot gas in the ISM. For example, for solar neighborhood-like gas surface densities and star formation rates, early SILCC simulations find most of the volume is filled with hot gas ($f_\mathrm{v, hot} > 0.5$) \citep{Walch2015}, which is consistent with the \cite{McKee1977} prediction, while TIGRESS find $f_\mathrm{v,hot} \sim 0.4$ and $f_\mathrm{v, warm}\sim 0.55$ \citep{Kim2017, Kado-Fong2020}. These latter values are more consistent with later SILCC simulations \citep{Peters2016, Rathjen2023}, but those works link the effect to the inclusion of radiative feedback, which is absent from the earlier TIGRESS models. In fact, in a more recent version of the TIGRESS simulations that includes radiation, \cite{Kim2023} find $f_\mathrm{v,warm} \sim 0.6$ and $f_\mathrm{v, hot} \sim 0.15$. Thus there appears to be a general trend that the inclusion of radiation tends to increase the warm gas volume filling fraction and decrease $f_\mathrm{v,hot}$, but exact fractions still clearly depend on details of the simulation implementation.

\paragraph{Wind structure and cosmic rays} Although most of the points above are largely agreed upon, recent efforts to include cosmic ray feedback in these simulations have called into question even many of the qualitative results discussed above. In particular, simulations that include cosmic ray injection as part of their supernova feedback tend to drive outflows with qualitatively different multiphase structure. The outflows in these simulations are denser, colder, and slower than outflows that are purely thermally driven, and above $\sim~1$ kpc, cosmic ray pressure support can dominate over thermal pressure, such that cosmic rays become the primary acceleration mechanism for warm gas \citep{Girichidis2016a, Simpson2016, Armillotta2021}. This effect can increase the total amount of mass being driven out in winds by a factor of $\sim 2$ \citep{Girichidis2018}, by converting a fountain flow (primarily driven by thermal supernova pressure) into an outflow even at solar neighborhood conditions.

In addition, at higher gas surface densities ($\Sigma_\mathrm{gas} \sim 100 \mathrm{M}_\odot\,\mathrm{kpc}^{-2}$), the inclusion of cosmic rays qualitatively changes outflows from primarily hot (at $z > 2$ kpc), to multiphase including cold gas \citep{Rathjen2023}. Although on the whole the warm gas distributions in CR-driven outflows are predicted to be smoother and slower, the continuous acceleration due to the CR-pressure gradient can lead to  velocities of up to hundreds of km/s at higher $z$ even in Milky Way-like potentials \citep{Farber2018}, which may prove to be a key observable diagnostic for the importance of cosmic ray feedback, a point we will return to in the following section. That said, the details of cosmic ray transport in these subgrid models is still highly uncertain and numerical implementations are varied, and it remains to be seen whether consensus will be reached on their impact in driving outflows across a range of galaxy masses \citep[see, e.g., reviews by][]{Hanasz2021,Ruszkowski2023}.

\subsection{Isolated galaxies}
\label{sec:disks}

In some ways, the simulations described in \S~\ref{sec:ISMpatch} represent the sweet spot between resolving relevant physical processes in the ISM, and capturing integrated features of galaxies relevant on larger scales. On the other hand, because these simulations are typically carried out in boxes with periodic $x$- and $y$-boundaries, they are unable to fully capture the global geometry of galaxies. This can play a role in e.g. the contribution of galactic shear or inflows in generating ISM turbulence, as well as the structure of multiphase outflows. In particular, periodic-box simulations cannot capture the opening of streamlines as winds expand into the halo, which limits their capability to capture the conversion of thermal to kinetic energy in the hot phase (and thus, their ability to accurately model the subsonic to supersonic transition in large-scale supersonic winds) \citep[see][]{Martizzi2016}.

In this section, we discuss a largely complimentary group of simulations that have been carried out using isolated global galaxy geometries. While these simulations must typically sacrifice physical resolution relative to the patch simulations in order to capture the extent of galaxy disks and the inner halo, they play a key role in understanding the link between the small scale mixing properties that are well captured in the ISM-scale simulations described in the previous section, and the halo-scale simulations that will be described in \S~\ref{sec:cosmo}. An important focus in these isolated galaxy simulations in recent years has been on capturing the multiphase structure in the ISM, rather than relying on a pressure floor or single-phase ISM model as has commonly been used in larger-scale cosmological simulations. Thanks in part to massive increases in computational resources, many of these global galaxy simulations now achieve resolutions sufficient to capture important physical processes that set the overall multiphase structure of the ISM (such as supernova explosions) at a scale similar to the periodic-box simulations discussed in the previous section, allowing for interesting comparisons between the two geometries. However, it is important to note that the vast majority of the simulations discussed in this section employ adaptive resolution methods of some kind (which can affect, e.g., the mixing processes in a multiphase wind; see ~\S~\ref{sec:computational_challenges}). 

Our focus here will primarily be on the structure of multiphase outflows in these simulations, as many of the conclusions regarding the structure of the ISM from the previous section hold here. In particular, global galaxy simulations also find that the volume filling fraction of the hot phase in the ISM depends on the local star formation rate (and therefore supernova rate), and tends to increase in higher gas surface density regions (i.e. smaller radii in a MW-like galaxy) \citep[e.g.][]{Bieri2023}. However, as in the ISM patch simulations, the volume-filling fractions of different phases are not converged across simulations, and the effect of numerical resolution on these fractions is not well-understood; depending on the subgrid prescriptions employed, higher resolution can result in more hot gas \citep{Marinacci2019} or less hot gas \citep{Kim2018} in the ISM. As in the periodic box simulations, global galaxy simulations with a resolved three-phase ISM tend to find that the cold phase has the smallest volume filling fractions (10 - 15\%), but makes up most of the mass (85 - 90\%) \citep{Bieri2023}. However, incorporating realistic stellar radiation can dramatically increase the amount of gas in the warm phase relative to the cold phase \citep{Kannan2020}, and significantly impact the efficacy of supernovae at launching outflows \citep{Smith2021}. In addition, some work demonstrates that the inclusion of runaway stars can significantly increase both the mass and volume filling fractions of the hot phase \citep{Steinwandel2023}, leading to more efficient driving of outflows and significantly higher energy loading, though others find little-to-no impact from these supernovae \citep{Andersson2023}.

\subsubsection{Outflow kinematics}
\label{sec:outflow_kinematics}

Outflows are a key focus of global galaxy simulations with resolved multiphase structure. This is in part because outflows are observed to be ubiquitously multiphase (see Section \ref{sec:where_multiphase_gas_exists}), which makes them difficult to fully characterize even for a single galaxy without comprehensive observations from a suite of telescopes. Isolated galaxy simulations also fill an important niche between ISM patch simulations and cosmological simulations — they are high enough resolution to be able to drive outflows self-consistently from a multiphase ISM, but are also large enough to capture the evolution in outflows as they move away from the galaxy. Due to computational expense, early simulations in this area tended to focus on small boxes ($\sim1\,\mathrm{kpc}^3$) and/or dwarf galaxies, but more recent works have scaled up to include Milky-Way mass systems and typically extend many kpc into the halo (though rarely at comparable resolution).

Using the mass- and energy-loading factors defined in the previous section, a simple, spherically-symmetric model of a thermally-driven supersonic outflow can constructed \citep{Chevalier1985}. Neglecting gravity, radiative cooling, conduction, etc. and defining the mass and energy injection region, $R$, a self-similar solution can be derived\footnote{ More recent analytic models have included the effects of gravity, radiative cooling, etc. \citep[e.g.][]{Wang1995, Bustard2015, Thompson2015}. See \cite{Thompson2024} for a recent review.}. Assuming constant mass and energy injection within $R$, the outflow will transition from subsonic to supersonic at the edge of the injection region\footnote{This transition can only occur in a three dimensional geometry that allows for expanding streamlines, not in the plane-parallel geometries employed in ISM patch simulations. This is an important reason to compare the results of the simulations discussed in Section \ref{sec:ISMpatch} to those discussed in this Section.}, where the characteristic temperature and velocity will be:

\begin{equation}
T \sim 10^7\,\mathrm{K}\left(\frac{\eta_E}{\eta_M}\right), \quad v \sim 500\,\mathrm{km}\,\mathrm{s}^{-1}\left(\frac{\eta_E}{\eta_M}\right)^{1/2},
\end{equation}
assuming one $10^{51}$ erg supernova per $100\,M_\odot$ of star formation. Outside the sonic point the gas will continue to accelerate, reaching an asymptotic velocity
\begin{equation}
v_\infty = \sqrt{2 \dot{E}/\dot{M}} \sim1000\,\mathrm{km}\,\mathrm{s}^{-1}\sqrt{\eta_E/\eta_M}
\end{equation}
\citep[see][]{Thompson2024}. These values are roughly in line with those measured in the global galaxy simulations discussed below, though the interaction of the hot driving phase in this wind model with cooler phases can significantly alter the properties of both \citep{Schneider2020}.

\subsubsection{Dwarf galaxy models}
\label{sec:dwarf_galaxies}

In some of the earliest superbubble simulations employing a global geometry and a multiphase ISM, \cite{Cooper2008} showed that multiphase outflows could be seeded by superheated supernova ejecta propagating through low density channels in the ISM, expanding outward in a thermal-pressure driven wind, and dragging warm ISM gas along. Although these simulations were carried out in a small volume for a relatively short time frame, many of the conclusions held in later works as well. In simulations with individual supernova explosions seeded at density peaks in the ISM, \cite{Fielding2017g} demonstrated that winds driven by non-clustered supernovae have too little mass and energy in all phases, emphasizing the importance of supernova clustering in driving outflows (see also Section \ref{sec:superbubbles}).

A key goal of many of these dwarf galaxy simulations was to develop a resolution criteria for accurately capturing the generation of winds. Using SPH simulations of a $10^{10}\,\mathrm{M}_\odot$ mass dwarf, \cite{Hu2019} showed that wind properties converge when the cooling masses of individual SNe remnants are resolved, which requires a mass resolution of approximately $5\,M_\odot$ in a mass-based refinement scheme. When SNe are resolved, the exact subgrid energy injection scheme does not affect the resulting wind structure, but injecting only terminal momentum cannot capture winds fully because it misses the hot driving phase \citep{Hu2019}. Similarly high resolution ($< 20\,M_\odot$ per volume element) was shown to be required in other adaptive refinement schemes to achieve realistic wind mass loading \citep{SmithM2018}. However, it is not clear that simulations have converged on mass loading rates in these dwarf galaxy outflows; some find relatively large factors ($\eta_M\sim10$) close to the galaxy that drop off significantly ($\eta_M\sim~1$) further out \citep{Emerick2018, Hu2019}, while others find significantly higher rates ($\eta_M \sim 10 - 100$) and very high energy loading ($\eta_E\sim1$)\citep{Andersson2023}. These effects may be partly due to different methods of measurement and time averaging, but it is not clear that these reasons alone can erase the discrepancies. In addition, in small galaxies, burstiness of the star formation rates may play a significant role; in non-equilibrium systems this may lead to very large instantaneous outflow rates which can obscure both simulation measures and observational estimates \citep{Martizzi2020}.

On slightly larger scales in an LMC-mass model ($M\sim10^{11}\,\mathrm{M}_\odot$), \cite{Steinwandel2024} find mass loading factors of $\eta_M\sim 1$ at close distances (1 kpc) and $\eta_M\sim 0.1$ farther away (10 kpc), while the energy loading is only $\eta_E\sim0.01$, with the bulk of the mass still transported in the warm phase, but comparable energy in the warm and hot phases. This is similar to what is found in \cite{Schneider2024} for a similar mass galaxy; these simulations are the only ones to report comparable energy loading in both the warm and hot phases, indicating that sufficiently high resolution is necessary to properly capture the energy transfer -- \cite{Steinwandel2024} employ a constant mass resolution of $4\,\mathrm{M}_\odot$ in a meshless-finite-mass scheme, while \cite{Schneider2024} have a constant spatial resolution of $\Delta x\approx 5\,\mathrm{pc}$ in a fixed-grid model.

In principle, large-scale thermal instability could explain the fast-moving warm gas observed in outflows if enough mass can be mixed into the hot phase. The theoretical argument is similar to that of cloud survival discussed in Section \ref{sec:cc} — if the cooling time is short relative to the dynamical time of the outflow, the hot phase can cool \citep{Wang1995,Thompson2015}. In particular, one can derive a ``cooling radius'' where the advection time equals the cooling time that depends strongly on the mass and energy loading factors in the hot phase, $r_\mathrm{cool} \propto \eta_\mathrm{E}^{2.13} / \eta_\mathrm{M}^{2.92}$ \citep{Thompson2015}. While this large-scale cooling was demonstrated to occur in simulations with artificially-increased values of $\eta_\mathrm{M, hot}$ \citep{Schneider2018b, Nguyen2024}, in more realistic models it is difficult to mix sufficient gas into the hot phase on short enough timescales to trigger bulk cooling — as noted in many simulations described above, $\eta_\mathrm{M, hot}$ is rarely observed to reach values higher than $\sim0.2$, while the critical value required to trigger large-scale thermal instability on scales comparable to the galaxy is $\sim1$ \citep[see][Eq. 14]{Thompson2024}. In addition, although the hot phase should be highly metal enriched via supernova ejecta (and thus should cool relatively efficiently), more efficient metal diffusion is shown decrease the overall cooling rate in the hot phase \citep{Steinwandel2024b}. Thus it appears likely that in the majority of systems, localized mixing-induced cooling (as described in Section \ref{sec:cc}) is the more common way to convert hot-phase gas to fast-moving warm outflows.

Unlike in idealized cloud-crushing simulations, however, the process of warm gas acceleration through mixing and momentum transfer happens globally in outflows, and can transport warm gas well out into the halo. In high-resolution global simulations, the measured acceleration of the warm gas is completely consistent with the majority of the momentum transfer happening through mixing \citep{Schneider2020} in agreement with single-cloud simulations (see ~\S~\ref{sec:cc}). The properties of the hot phase are also affected by mixing, such that the densities and temperatures are higher than are predicted in a pure adiabatic expansion model, and the velocities are slower. This mechanism can explain the observed soft X-ray luminosities of local starburst galaxies \citep{Schneider2018a, Nguyen2021, Schneider2024}. The geometry of the starburst can also play a role in the phase structure of the outflows, particularly in the hot phase. More centrally-concentrated bursts, or even ring structures, lead to hotter, faster outflows, even approaching speeds more commonly associated with AGN-driven winds \citep{Nguyen2022, Schneider2024}.

Although many of these simulations represent the highest resolution global galaxy models to-date, it is still not clear that all of the above results are converged (given that many of the characteristic length-scales of multiphase systems discussed in the previous sections are generally still unresolved). For example, using a modified version of an adaptive-mesh-refinement scheme, \cite{Rey2024} demonstrate that increasing the resolution in low density gas (by refining on the cooling length) can affect the hot phase significantly -- outflows reach higher temperatures and stay hotter \citep[see also][]{Smith2024}, and they note up to a 5-fold increase in the energy loading. Increasing resolution also increases the amount of colder denser gas in the outflow, a similar trend to that noted in the ISM phase structure in stratified disk simulations.

\subsubsection{Milky-Way-mass models}

With the additional computing power available in recent years, there have been efforts to extend the realistic, 3-phase ISM models previously explored primarily in stratified disk simulations and dwarf galaxy models to larger global galaxy models. Several groups have implemented different versions of these models, and find largely consistent results regarding outflows. In particular, MW-mass simulations find warm outflows near the disk with mass-loading factors of $\eta_\mathrm{M, warm} \sim 1$, but these are largely fountain flows on a $\sim100$ Myr timescale \citep{Marinacci2019, Bieri2023}. This is due to the impact of the increased gravitational potential \citep{Tanner2020}, which is consistent with the tall-box simulations that find much steeper relationships between galaxy mass (or relatedly, gas surface density) and warm gas mass loading than with hot gas mass loading. Also consistent with the stratified disk models, the majority of the outflowing mass is found in the warm phase, while the majority of the energy is carried in the hot phase \citep{Armillotta2019, Bieri2023}, but energy loading factors are typically very low.

\begin{figure}
\centering
\includegraphics[width=\linewidth]{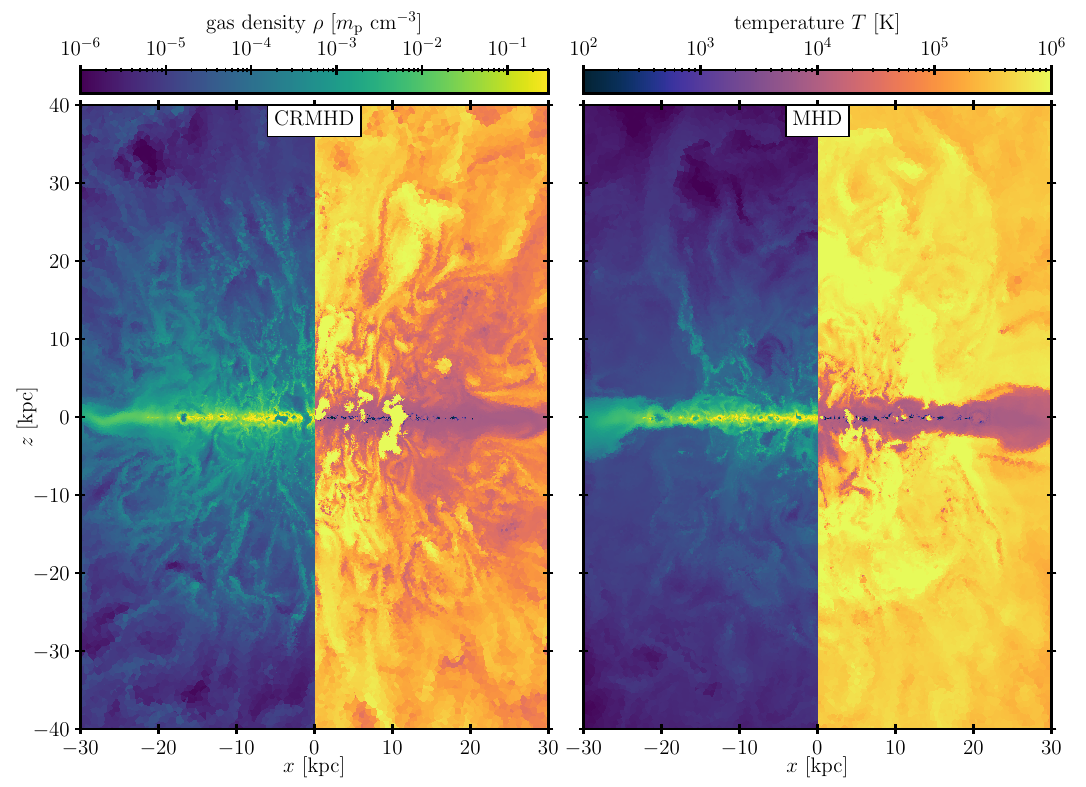}
\label{fig:CR_outflow}
\caption{Density and temperature slices that show differences in the multiphase structure of a simulated outflow in a isolated Milky Way-mass galaxy simulation with (left) and without (right) cosmic ray feedback. Adding cosmic rays increases the total amount of cool, dense gas outside of the galaxy, and lowers the overall temperature of the outflow. Figure from \cite{Thomas2024}.}
\end{figure}

As in the tall-box simulations, however, the inclusion of cosmic rays as a form of supernova feedback adjusts this picture somewhat \citep{Uhlig2012, Salem2014a, Butsky2018}. Particularly for more massive galaxies where the galactic potential limits the reach of warm thermally-driven outflows, cosmic-ray driven winds can look quite different. By coupling to the warm gas outside the disk, they can boost overall mass loading factors by factors of $\sim2 - 10$ (depending on the chosen parameters and included transport mechanisms), increasing the total amount of outflowing warm and intermediate-temperature gas (see Figure \ref{fig:CR_outflow}). These outflows are cosmic-ray pressure dominated outside of the galaxy ($z \gtrsim 1$ kpc), and can change the picture of a warm fountain flow to a slow, steady outflow \citep{Thomas2024}.

Similarly, \cite{Girichidis2024} show that adding cosmic rays impacts not only the outflow properties, but also the larger-scale halo properties (see also \citealp{Ji2020a}), including in the case where there are no significant outflows. For lower-mass $M_{\rm halo} \sim 10^{11}\,{\rm M}_\odot$ galaxies, they find that cosmic ray pressure support leads to a quasi-steady hydrostatic halo at temperatures $T\sim 10^5\,\mathrm{K}$ that would be unstable in a halo with purely thermal support. For larger, Milky-Way-mass galaxies, they find that the effect of cosmic rays is primarily to significantly puff up the disk-halo interface, leading to large clouds of intermediate temperature gas in the inner halo at $z$ heights of 10-20 kpc that is absent in models without cosmic ray feedback.

\paragraph{Summary regarding loading factors}

\begin{figure}
\centering
\includegraphics[width=0.49\linewidth]{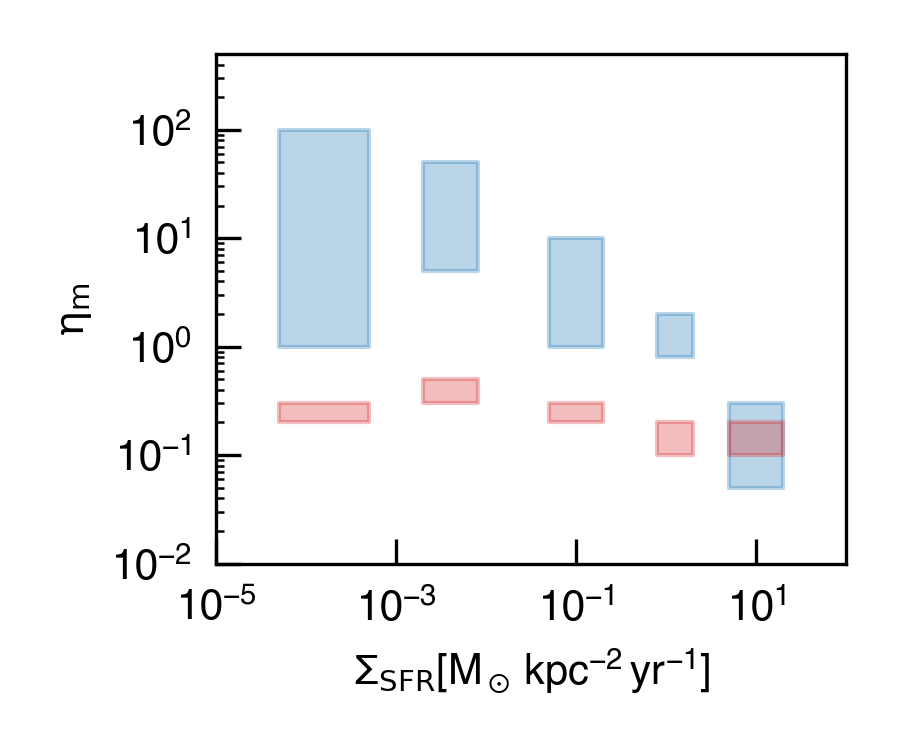}
\includegraphics[width=0.49\linewidth]{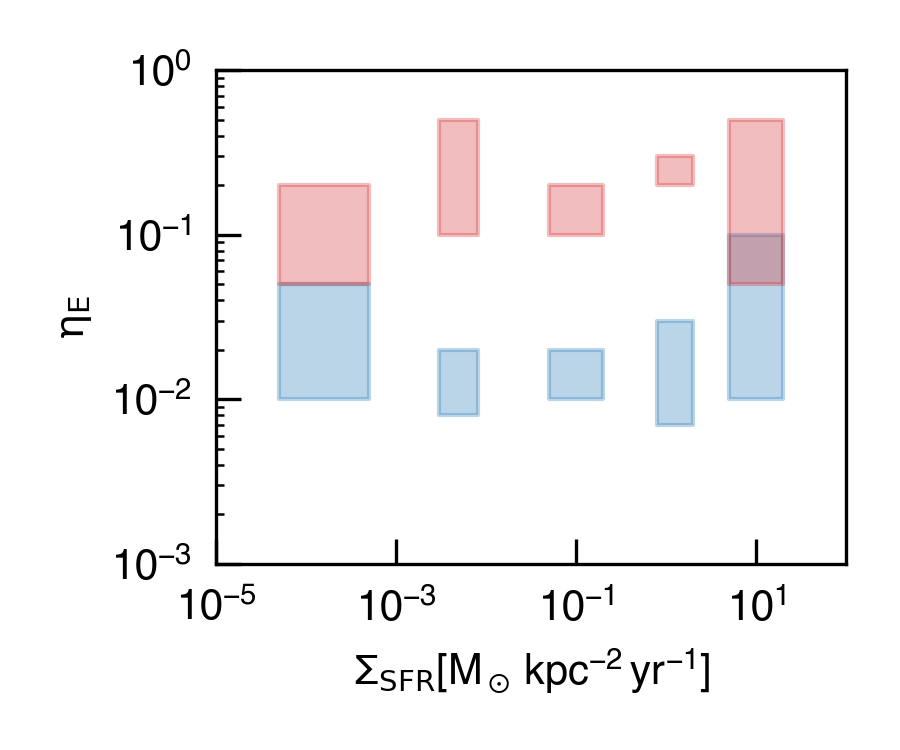}
\caption{Mass (left) and energy (right) loading rates as a function of star formation rate surface density for warm (blue) and hot (red) gas measured from a variety of different simulations discussed in Sections \ref{sec:ISMpatch} and \ref{sec:disks}. Ranges are approximate.}
\label{fig:loading_factors}
\end{figure}

Although these galaxy-scale simulations have yet to fully converge on exact results, we can attempt to summarize the picture for measured mass and energy output in both the tall box simulations discussed in Section \ref{sec:ISMpatch} and the isolated galaxy simulations discussed in \ref{sec:disks}. In Figure \ref{fig:loading_factors} we plot the range of mass and energy loading factors found in these simulations as a function of star formation rate surface density ($\Sigma_{\rm SFR}$). As mentioned previously, quantifying these loading factors with a single number is difficult even for a single simulation, since the results will depend on where the rates are measured, whether one uses spherical versus planar geometry, what phase is being considered, etc. Nevertheless, some conclusions can be drawn. Overall, the cool gas mass loading decreases as $\Sigma_{\rm SFR}$ increases, indicating that mass removal is more important for lower-mass galaxies. Hot gas mass loading is relatively constant, with $\eta_{\rm m, hot} \sim 0.1$ across a wide range. Conversely, hot gas energy loading increases as a function of $\Sigma_{\rm SFR}$, pointing towards more effective breakout and a larger role for preventative feedback via high specific energy outflows in more massive or starbursting systems.

\subsection{Halo scale}
\label{sec:cosmo}

As discussed in Section \ref{sec:where_multiphase_gas_exists}, the circumgalactic medium (CGM) and intracluster medium (ICM) are observed to host multiphase gas. However, the large scales of these systems (typically hundreds of kpc) as compared to the scales that must be resolved in order to fully capture the multiphase structure in a simulation are daunting, and in many cases completely computationally infeasible; this challenge led to many of the small-scale idealized simulations discussed in previous sections. Nevertheless, to gain a complete understanding of the gas in galaxy and cluster halos, their cosmological environment cannot be ignored, as it contributes significantly to setting the overall temperature, pressure, and velocity distributions of the gas \citep[see e.g.][]{Fielding2020b}. Recognizing this problem, many groups in recent years have begun to push on the resolution that can be achieved in simulations that capture galaxy outskirts within a cosmological context. In this section we focus on three approaches to modeling multiphase gas in galaxy and cluster halos: idealized, traditional cosmological zooms, and enhanced CGM resolution zooms.

\subsubsection{Halo scale idealized (stratified)}
\label{sec:haloscale}

In Section \ref{sec:turbulent} we discussed results from many simulations that focused on thermal instability in halo-like conditions. Here we focus on results from halo-scale simulations that include stratification of the background due to the gravitational potential of the central galaxy or cluster. In contrast with the following sections, the simulations discussed here do not attempt to model in detail (or at all) the processes occurring within the central galaxy, though some include it as a source / sink term at the center of the halo. Most of these simulations take as their initial conditions a halo-scale box in hydrostatic equilibrium, which is then perturbed via a variety of mechanisms (cosmological accretion, galaxy winds, turbulent driving, etc.) and examine the resulting changes to the atmosphere.

In the standard picture of galaxy formation theory, halo gas is expected to be heated to the virial temperature behind an accretion shock, with the virialized gas in hydrostatic equilibrium with the dark matter halo potential \citep{Rees1977}. For massive halos ($M_\mathrm{halo} > 10^{11.5}\, M_\odot$), this temperature is high ($T \gtrsim 10^{5.5}$ K) and the gas cools on a timescale longer than the free-fall time; in less massive systems, the halo gas can cool quickly. In early simulations of this process, \cite{Birnboim2003} showed that in low-mass halos, inflowing gas does not form an accretion shock when $t_\mathrm{cool}$ for the post-shock gas is less than $t_\mathrm{ff}$. This led to the paradigm of ``hot mode” ($t_{\rm cool} \gg t_{\rm ff}$) versus ``cold mode” ($t_{\rm cool}\ll t_{\rm ff}$) accretion for galaxies depending on the average state of their halo gas, though as noted in \S \ref{sec:turbulent}, simulations of thermal instability which account for a feedback heating term indicate that precipitation and cooling of over densities can lead to accretion of cold gas even in hot halos.

\cite{Fielding2017a} confirmed this picture numerically in the presence of stellar feedback, showing that the critical halo mass to form a hot, thermally-supported halo is not significantly affected by galaxy winds. However, the multiphase structure of lower-mass halos below the critical threshold does depend sensitively on the wind properties; high mass-loading ($\eta_\mathrm{m} > 1$) leads to a predominately low-temperature CGM supported by turbulence, while high specific-energy winds produce halos with appreciable intermediate temperature gas ($T \sim 10^5 - 10^6$ K).

Within the hot mode accretion regime, a special steady-state solution exists. Because the cooling time in the inner regions is shorter, gas can cool on timescales less than the Hubble time, $t_{\rm Hubble}$. At radii where $t_{\rm ff} < t_{\rm cool} < t_{\rm Hubble}$, compressional heating balances radiative losses, leading to $t_{\rm flow} \approx t_{\rm cool}$ where $t_{\rm flow} \equiv r/v_r$, i.e., a steady-state and subsonic `cooling flow' solution exists \citep{Mathews1978, Fabian1984, Fabian1994}. While observations rule out the widespread existence of such cooling flows in galaxy clusters \citep[e.g.][]{Peterson2003,McDonald2018}, \citet{Stern2019} adapt this solution for the CGM around Milky Way mass galaxies, where the observational verdict of the presence of cooling flows is still outstanding. Furthermore, \cite{Stern2020} showed that in dwarf mass halos and at inner CGM radii of $L^*$ galaxies, $t_{\rm cool}$ of the hot gas is shorter than $t_{\rm ff}$ and thus compressive heating is not sufficiently rapid to balance radiative heating. The hot inflow then experiences a large-scale thermal instability qualitatively similar to that discussed for outflows in $\S \ref{sec:dwarf_galaxies}$. This large-scale instability is seen in the FIRE cosmological zoom simulations \citep{Stern2021, Gurvich2023, Kakoly2025}.

Subsequent numerical work has included a variety of additional processes which may affect this picture, including magnetic fields, halo rotation, cosmic ray pressure, etc. Among the results that has emerged is the importance of rotation, which provides additional pressure support to hot halos and also enhances mixing, which can lead to increased cooling and condensation \citep{Buie2022}. \cite{Stern2024} derive an analytic solution (confirmed using idealized simulations) that models a subsonic accretion flow (akin to the ones discussed above) from a rotating hot halo that cools at the disk-halo interface, which may be the dominant form of accretion in $L^*$ galaxies \citep[see also][]{Hafen2022}. This model of a steadily inflowing hot halo contrasts with the typical thermal balance models invoked in many thermal instability studies \citep[e.g.][]{Sharma2012, Voit2017}, which assume a static hot halo where heating due to feedback balances radiative losses and accretion is primarily driven by precipitation in regions where $t_\mathrm{cool} < t_\mathrm{ff}$ locally (see \S~\ref{sec:turbulent}).

On the larger scales of the intracluster medium, the prevailing picture is one where gas is heated to the viral temperature via shocks (as above), but the higher pressures in the inner regions lead to cooling times that are much less than the Hubble time. As discussed above, these `cool core’ clusters require that a heating source exist to prevent the development of cooling flows and match X-ray observations. One possible source of heating is conduction from the outer regions — \cite{Ruszkowski2010} showed using 3D MHD simulations that low levels of subsonic turbulence can randomize the magnetic field in clusters and enhance the conductive heat flux sufficiently to avoid the cooling catastrophe in low density clusters. However, conduction alone is insufficient to prevent the necessary cooling in many systems. In the inner regions, the heating source is typically assumed to be feedback via jets from an AGN at the cluster center \citep[e.g.][]{Bruggen2009}, though turbulence driven by minor mergers and the motions of cluster galaxies may also play a role. 

Using 3D AMR simulations, several groups showed how these jets can generate turbulence that creates local conditions in which $t_\mathrm{cool} / t_\mathrm{ff} < 10$ (see \S~\ref{sec:turbulent} for a discussion of this criterion), driving thermal instability and the precipitation of cool gas, while the atmosphere as a whole remains stable \citep{Gaspari2011, Gaspari2012, Li2014, Li2015a}. Given the high pressures in the inner regions of clusters, this gas can cool below the typical $T\sim 10^4$ K equilibrium in $M_\mathrm{halo} \sim 10^{12}\,M_\odot$ halos; indeed the density peaks in condensing regions can reach $T < 50$ K and even form molecular gas, with kinematics inherited from the X-ray plasma via the turbulent cascade, and consistent with observations \citep{Gaspari2017}. Critically, this cool gas then `rains down’ on the AGN, feeding the supermassive black hole and raising the heating rate and ICM entropy until $t_\mathrm{cool} / t_\mathrm{ff} > 10$ globally and the system stabilizes \citep{Gaspari2011, Gaspari2012, Li2015a}. This so-called `chaotic cold accretion’ differs significantly from classic accretion models such as Bondi, the cooling flow solution mentioned above, and the thin-disk approximation, and can result in accretion rates that are orders of magnitude larger than the time-steady solutions \citep{Gaspari2013}.

\subsubsection{Cosmological zooms}

While the models discussed above have provided valuable insight into the overall phase structure of galaxy halos, their simplified setups mean that they nevertheless cannot capture the full cosmological context in which halos evolve. In contrast, zoom-in cosmological simulations have long been used to increase the resolution of a single galaxy while still capturing the effects of its cosmological environment \citep{Katz1993}. In brief, to create a zoom, a low-resolution simulation is carried out in a cosmological box and run for sufficient time to isolate the halo (or halos) of interest. These regions (or Lagrangian tracers) are then re-simulated at much higher resolution within the lower resolution box, allowing the majority of the computing power to be spent resolving properties in the halo of interest, while still capturing the bulk kinematics and structure of the cosmological background. While the resulting spatial (or mass) resolution is still typically several orders of magnitude larger than the scales in the stratified disk or isolated dwarf galaxy simulations discussed above\footnote{A typical gas mass resolution element in a modern zoom is $\sim 10^3\,\mathrm{M}_\odot$, corresponding to a spatial resolution of $\approx 75$ pc for gas with a typical warm ISM number density $n = 0.1\,{\rm cm}^{-3}$.}, it can be sufficient to resolve key features of galaxies, for example, the scale-height of the warm ISM in disks, or the bulk properties of giant molecular clouds.

\textbf{Outflows:} Like their higher-resolution counterparts, zoom simulations typically find multiphase structure in outflows, though it should be noted that their properties are usually measured at much larger distances from the central galaxy -- $0.25\, R_\mathrm{vir}$ is typical \citep{Muratov2015}. In addition, due to the lack of sufficent resolution in the ISM to capture the hot driving phase, these larger scale simulations commonly employ subgrid recipes for wind launching. Thus, quantitative comparisons with smaller scale work can be challenging, since e.g. the measured mass-loading factor can be significantly influenced by swept-up CGM material, and time averaging is necessary to account for the travel time of outflows relative to the instantaneous star formation rate. Nevertheless, the qualitative picture from zooms largely agrees with that of the stratified disk and isolated galaxy simulations. \cite{Pandya2021} find mass-loading factors that scale strongly with galaxy mass, ranging from $\sim100$ for low-mass dwarfs to $< 1$ for Milky-Way mass systems in the FIRE-2 simulation suite \citep{Hopkins2018}. Trends are also found between outflow temperature and galaxy mass, with most of the mass carried by the hot phase in high mass galaxies, and by the warm phase in dwarfs. While these trends are consistent with the results of higher resolution simulations discussed in Sections \ref{sec:ISMpatch} and \ref{sec:disks}, there are significant quantitative differences, for example, higher mass loading factors across all masses.

\textbf{CGM:} Consistent with larger-scale cosmological simulations, zooms find $T \sim 10^4$~K clumps of gas embedded within higher temperature halos \citep[e.g.][]{Oppenheimer2018, Esmerian2021, Appleby2021}. In principle, these clumps can explain the absorption-line observations of low ionization metal lines in the halos of galaxies, though in practice, the process of translating between these simulation data and the observables is not trivial, and there exists significant disagreement about whether the current generation of simulations does or does not reproduce key observables. This is likely due in part to a lack of convergence, particularly in the cool phase — as the resolution is increased, the median cloud masses and sizes decrease \citep[e.g.][]{Ramesh2023}. Simulations commonly find that these cool clouds are under-pressurized with respect to the background hot gas, which in some cases indicates the importance of non-thermal pressure support \citep[i.e. magnetic fields,][though note that their clouds are still under-pressurized when accounting for the total pressure]{Ramesh2023}. Other studies attribute these lower pressures to insufficient numerical resolution \citep{Oppenheimer2018}. In general, higher resolution models are better able to capture the generation of cool gas, because local differences relative to the average can significantly alter halo gas evolution; for example, density perturbations seeded by orbiting substructures can lead to rapid cooling and condensation of hot halo gas even when the median $t_\mathrm{cool}/t_\mathrm{ff}$ is higher than 10 or 20 \citep{Saeedzadeh2023}.

As is the case for smaller-scale models, the inclusion of cosmic rays in cosmological zooms is also found to alter the multiphase properties the CGM \citep[e.g.][]{Salem2014b, Salem2016}. However, it is not clear that simulations have reached consensus on the magnitude of these effects. In Milky-Way-mass zooms, \cite{Buck2020} find that the inclusion of CR feedback leads to halos that have a cooler and smoother gas distribution, with the cosmic rays providing additional pressure support that stabilizes the gas. They find halos that can have large volume-filling regions at the (normally) thermally unstable peak of the cooling curve around $T\sim 10^5$ K. \cite{Ji2020a} find similar effects in the low redshift Milky-Way mass models of the FIRE-2 simulation suite, but tend to find even cooler gas ($T\sim 10^4 - 10^5$ K) with even higher levels of CR pressure support, and find that the preventative nature of CR feedback (which keeps this cool gas from raining down onto the galaxy as it otherwise would) can reduce star formation rates in these systems \citep{Hopkins2020}.

\subsubsection{Enhanced resolution CGM}
\label{sec:ehanced_resolution_cgm}

\begin{figure}
\centering
\includegraphics[width=\linewidth]{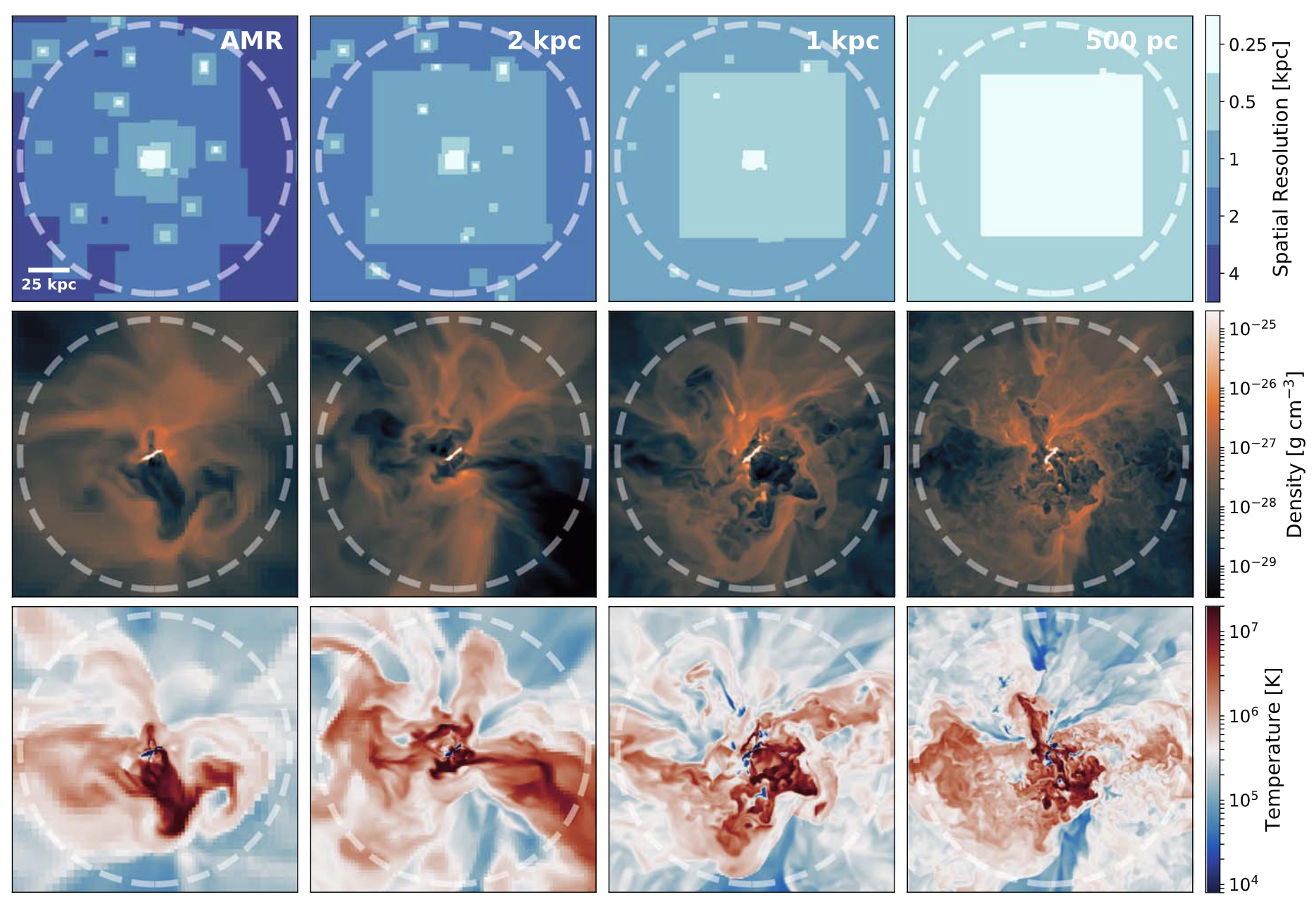}
\caption{Example of the effect of `enhanced halo resolution' on cool gas in a traditional zoom-in AMR model of an $L^*$ galaxy. Figure from \cite{Hummels2019}.}
\label{fig:EHR}
\end{figure}

Recognizing the need for higher resolution in halos, a number of recent models have focused on explicitly improving the resolution in the outskirts of galaxies. Because they are designed to track and refine based on mass or density, increasing the resolution in a standard particle or AMR cosmological simulation will result in increased spatial fidelity in the ISM of galaxies; after all, the gas in the ISM is denser than that in the CGM by orders of magnitude. However, it was recognized that by modifying the refinement criteria to focus explicitly on the gas outside of the ISM, models could be produced that increased the mass and spatial resolution in the CGM by large factors with relatively little additional computational expense \citep{vandeVoort2019, Hummels2019, Suresh2019, Peeples2019}. These ``super-Lagrangian” or ``enhanced halo resolution” methods can achieve spatial resolution in the CGM of better than 100 pc for cool gas\footnote{In this subsection we will use ``cool" to mean $T\sim10^4$ K, in order to maintain consistency with the literature.}, which is sufficient to start accurately capturing the evolution of kpc-scale clouds (see Figure~\ref{fig:EHR}). The exact method of refinement varies, often depending on the underlying numerical method. Some simulations employ a fixed spatial resolution in the CGM, targeting $\sim 0.5$ kpc scales regardless of gas phase \citep{Hummels2019, Peeples2019}. This approach leads to the best resolution of the lowest density gas, making it ideal for studies of e.g. outflow-driven turbulence in the hot phase gas \citep{Lochhaas2023}. Others use a fixed mass resolution that is smaller in the halo than in the galaxy ISM \citep{Suresh2019, Ramesh2024}, which leads to much better spatial resolution of the cool phase gas. Still others employ some combination, requiring a fixed spatial target while also employing a mass-resolving approach \citep{vandeVoort2019}.

Some results of these studies are consistent across all simulations and methods. For example, as CGM resolution is enhanced, cool gas clouds are resolved at progressively smaller scales, and the lower mass (or size) limit of these structures is not converged \citep[see also][who showed this effect using a traditional refinement method]{Faucher-Giguere2010}. This increase in small scale structure tends to increase the covering fraction of cool phase gas, changing observables like the HI column density around L* galaxies \citep{VandeVoort2018}. This effect tends to be most pronounced when the ``original” simulation CGM resolution is lower, i.e. larger differences are found when the spatial resolution of a typical cool gas element is decreased from 10 kpc to 1 kpc than from 1 kpc to 100 pc \citep[compare e.g.][]{Hummels2019, Suresh2019}. On the other hand, simulations disagree about whether the total mass in the cool phase changes as a function of enhanced halo resolution; some studies find consistent bulk properties and average radial profiles \citep{Suresh2019, Ramesh2024}, while others find significant increases in the cool phase mass \citep{Hummels2019}.

Another focus of these simulations is on untangling the origin of cool phase gas in the CGM. In principle, enhanced resolution should impact the evolution of the cool phase, for example by enhancing the seeding of condensation by better resolving existing cool structures (from satellites or winds) as well as decreasing artificial mixing of cool clouds \citep{Hummels2019}. Enhanced resolution should also better capture the potential shattering of over densities as they cool (see Section \ref{sec:turbulent}). Using particle-tracking techniques, some groups find that winds contribute significantly to the production of cool gas in the CGM, either by seeding it directly or by triggering cooling of existing halo gas \citep{Hafen2019, Suresh2019, Ramesh2024}.

\section{Summary and Outlook}
Before turning to open challenges and future directions, it is useful to briefly
recap the range of physical scales and numerical approaches discussed throughout
this review (and depicted in Fig.~\ref{fig:simulations}). Moving from the smallest scales to the largest, we began with idealized, small-scale simulations that isolate individual physical processes governing multiphase gas, including turbulent mixing layers, cloud-wind (and stream-background) interactions, thermal instability and turbulent boxes. These setups, typically spanning sub-parsec to tens-of-parsec scales, are designed to achieve high spatial and temporal resolution and enable controlled parameter studies, giving detailed insight into mixing, mass and momentum transfer, phase structure, and morphology -- as well as to assess the impact of other physics such as magnetic fields, conduction, viscosity and cosmic rays on key results. We then moved to progressively less idealized and larger-scale simulations, including supernova-driven bubbles, stratified disk patches, isolated galaxies, and cosmological zoom-in simulations, which capture the global context in which multiphase in galaxies and their environments gas forms and evolves, but necessarily rely on limited resolution and subgrid modeling to capture some physical effects.

A central theme emerging from this overview is that these different classes of simulations should not be viewed as independent or competing approaches, but as complementary components of a unified modeling framework.
Large-scale simulations set the boundary conditions, driving mechanisms, and relevant regions of parameter space for small-scale studies: they help in determining the typical pressures, densities, metallicities, turbulence levels, radiation fields, and kinematic environments in which multiphase interactions occur.
Conversely, small-scale simulations provide insight into the physical interpretation of processes operating in simulations of galaxy formation and evolution, and allow one to assess when and how unresolved microphysics matters for the macroscopic evolution seen in larger-scale simulations. Because of their idealized nature, they are ideal for pushing the development of analytic theory and testing the utility of simplified analytic approaches, for example in the context of mixing layer theory (\S \ref{sec:tmls}), cool gas survival criteria (\S \ref{sec:cc}), or the development of thermal instability in stratified backgrounds (\S \ref{sec:turbulent}). Furthermore, they can be used to test subgrid prescriptions, e.g., for feedback and star formation, but also for mixing, cooling, phase exchange, and momentum transfer in a multiphase medium (see \S~\ref{sec:subgrid}).

A key step in connecting these small-scale approaches to the larger simulations lies in testing the utility of the predictions made using idealized setutps in more realistic environments. For example, cloud-wind survival criteria can be tested on ensembles of clouds in realistic outflow environments in ISM patch simulations \citep{Tan2024a}, isolated galaxies \citep{Warren2025}, and the CGM \citep{Ramesh2025}. Similarly, thermal instability criteria can be tested using halo-scale models that include realistic sources of turbulence, such as AGN jets \citep{Gaspari2020} and satellite galaxies \citep{Hafen2022}.

Ultimately, however, simulations at all scales must be confronted with observations. Observations provide the ground truth -- they define which phases exist and where, how much mass they contain, how they move, and how they radiate, and therefore are the final constraint on any theoretical model. Large-scale simulations are essential for connecting predictions to global galaxy properties and observational selection effects, while small-scale simulations are critical for interpreting specific diagnostics, such as emission and absorption lines, covering fractions, and phase-resolved kinematics. While connecting simulations to observations in a realistic way can be challenging and will often require complex and detailed radiative transfer, the development of analyses that allow for apples-to-apples predictions in observational space is critical to actually constrain these model predictions. Progress in understanding multiphase gas in and around galaxies will therefore depend on closing the loop between simulations across scales and observations, using each to inform and refine the others.

\subsection{Technological advances}
\label{sec:tech_advances}

As pointed out in Section \ref{sec:computational_challenges}, a historic challenge in simulating multiphase systems is the necessity to faithfully capture a broad dynamic range in density, temperature, and length scales. While for some problems, adaptive refinement techniques or Langrangian simulation approaches are sufficient, for many of the problems discussed herein (mixing layers, turbulent boxes, outflows, etc.) mass-based refinement cannot capture the system well, because high resolution is needed to capture instabilities in gas which may be below the median density in the simulation. In addition, development of the underlying theory of multiphase systems requires robust numerical convergence studies, which are difficult (or impossible) to carry out with many adaptive refinement schemes. While clever modifications can be made to specific setups to reduce the overall computational cost (e.g. cloud-tracking in cloud-crushing simulations, or refinement based on cooling-length in CGM simulations, as discussed above), by-and-large these simulations require expensive, fixed resolution (or targeted static refinement) domains; a primary reason many of the earliest numerical models were carried out in 1 or 2 dimensions.

However, simulations of multiphase media also have an advantage over simulations of many other astrophysical systems, which is that in many cases, the required physics is purely local (complex as the governing equations may get). The fundamental equations of magnetohydrodynamics, optically-thin radiative cooling, and even many numerical models of radiative transfer and cosmic ray transport only require information from a local stencil, which makes such models a prime target for increased algorithmic parallelism. This convenient fact, combined with the vast increase of parallelism on modern supercomputers, has in large part led to the widely expanded field of simulation that we have attempted to describe in this review.

\begin{figure}
\centering
\includegraphics[width=\linewidth]{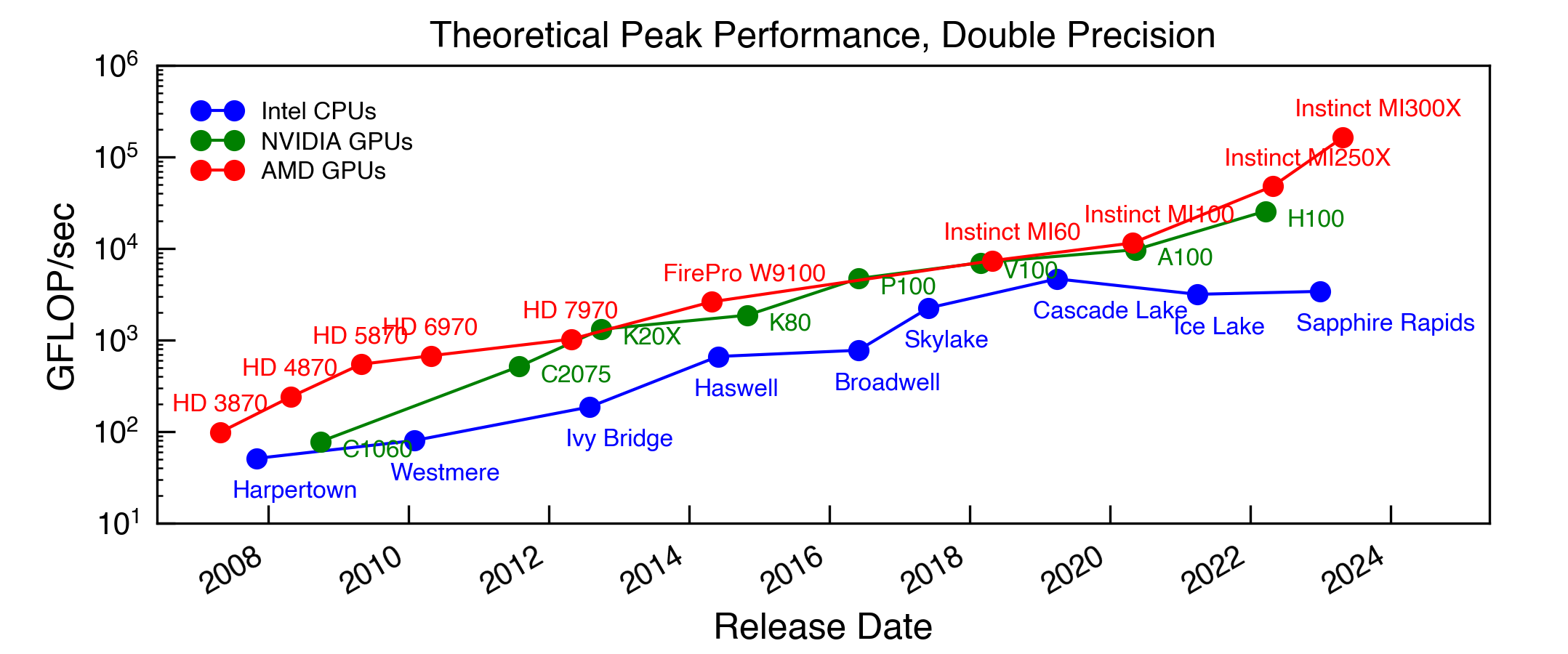}
\caption{A comparison of the theoretical peak performance for various generations of Intel CPUs, NVIDIA GPUs, and AMD GPUs. Note that in practice, most codes do not achieve even close to the peak performance of a given processor, nevertheless, this trend demonstrates the rationale behind the wide acceptance of GPUs in high performance computing.}
\label{fig:cpu_vs_gpu}
\end{figure}

For many decades, increased computational power primarily derived from increases in CPU-clock speeds, leading to very powerful single-core systems. Starting in the mid-2000’s, however, cooling requirements led to a plateau of clock speeds, such that increased computational power could only be derived from adding additional cores. This led to an early generation of parallelized astrophysics codes that could take advantage of multi-core, multi-node systems, e.g. Flash, \citep{Fryxell2000}, RAMSES, \citep{Teyssier2002}, Gadget \citep{Springel2005}, Athena \citep{Stone2008}, Piernik \citep{Hanasz2010a}, AREPO \citep{Springel2010}, and Enzo \citep{Bryan2014}). Since then, this trend towards increased hardware parallelism has exploded, particularly with the advent of graphics processing units (GPUs), which were originally designed for fast 2D pixel rendering. The GPU takes the parallel programming model to the extreme, with a single chip hosting thousands of individual, low-clock-speed cores, and devoting far more of the transistors to calculations, rather than control flow. This led to large gaps between the theoretical peak performance of the typical GPU versus the typical CPU (see Figure \ref{fig:cpu_vs_gpu}), and their subsequent adoption on many supercomputers. The advent of high-level programming models (i.e. CUDA) in the early 2010’s led to the subsequent adoption of GPUs in many scientific codes, transforming the HPC landscape in the following decade.

Early adoption of GPUs in astrophysics codes showed extreme promise, demonstrating that they could provide large speedups over similar CPU-based algorithms, for example in the GAMER \citep{Schive2010} and Cholla \citep{Schneider2015} codes \citep[see also][]{PortegiesZwart2007}, provided careful data management and that the GPUs were given sufficient work to do. Subsequent years have seen the adoption of GPUs by many additional groups, and virtually all of the world’s fastest supercomputers now employ GPU (or closely related) technology, including the first exascale systems\footnote{https://www.top500.org}. As a result, there now exists a wide array of highly parallelized hydrodynamics codes that are optimized for astrophysical applications and modern supercomputer architectures. Recent additions to this space include DISPATCH \citep{Nordlund2018}, Phantom \citep{Price2018}, QUOKKA \citep{Wibking2022}, Parthenon \citep{Grete2022}, SWIFT \citep{Schaller2024}, and AthenaK \citep{Stone2024}. These technological and algorithmic advances  have made extremely high resolution ($> 10$ billion resolution elements) hydrodynamics simulations feasible for the first time, and made parameter studies of lower resolution models cheap and accessible to a broad swath of modelers \citep[e.g.][]{Schneider2018a, Grete2023, Fielding2023}. These computational advances (including better / increased parallelism in CPU-based codes) have powered many of the results described herein. As more simulation codes take advantage of this technology and subsequent generations of GPUs continue to push the envelope of HPC systems, we will inevitably be able to directly simulate an increasingly broad range of scales.

However, challenges remain. In particular, even with continued increases in computing power, there are still $\sim$4 orders of magnitude between the resolutions currently feasible and what would be required to fully resolve the multiphase gas in e.g. a single Milky Way halo (see \S\ref{sec:computational_challenges}). In addition, adaptive time-stepping is often employed in simulations of many of the more realistic systems discussed in this review (the ISM, galactic winds, the CGM, etc.) in order to reduce computational expense. This approach can save huge amounts of compute time by allowing different cells / resolution elements to take different $\Delta t$, since the timestep in a non-adaptive method is set by the fastest signal speed in the domain, which may comprise only a small fraction of the volume. Indeed, many codes already employ adaptive time-stepping to great effect \citep[e.g.][]{Springel2010}. However, in a massively parallel code with adaptive time-stepping, efficiently utilizing the full power of the computer hardware can be nontrivial, since it requires that different resolution elements follow different control paths, making this method much more complex to load-balance across nodes and difficult to port to GPUs.

Another challenge is software portability. While most modern astrophysics software is built on a C++ or Fortran framework and can be compiled and run on virtually any CPU, GPU programming languages have yet to be standardized. Many early GPU codes adopted the CUDA programming language (developed and popularized by NVIDIA), but CUDA has the downside of being proprietary and only works on NVIDIA hardware. Other GPU manufacturers have developed their own similar programming languages (e.g AMD's HIP or Intel's SYCL), but this still leaves the problem of porting code up to the programmer. A recent path forward in this space is the emergence of C++ programming libraries like Kokkos, Raja, OpenACC, and OpenMP, which incorporate a flexible parallel programming framework that can be compiled for a variety of architectures, including both CPUs and GPUs. While these libraries do not currently match the sophistication or have as many features as the proprietary languages, they have the advantage of allowing software to be explicitly portable, and some version of these programming frameworks may ultimately become part of the C++ programming standard.

A final challenge is data management. Although it is now feasible to run simulations with tens of billions of resolution elements, saving this data for post-processing and analysis is often prohibitively expensive, as a single snapshot can comprise terabytes of storage space. Increasingly, the trend has been for compute power to outpace storage capabilities, and in particular, long-term data storage for these datasets is extremely costly. Thus, in-situ analysis methods must be developed, but these are often custom-built for the software being used to run the simulation, making them both time-consuming to develop and non-transferrable between codes. Increasingly modular software simulation frameworks that can be `plugged in' to a variety of codes may help with this issue \citep[e.g.][]{Grackle2017}, as well as centralized and public data hosting centers.

\subsection{Subgrid models}
\label{sec:subgrid}

As discussed in \S~\ref{sec:computational_challenges}, simulating multiphase gas flows in galaxy formation and evolution is a challenging problem because of its multi-scale nature. Some of the processes discussed earlier—such as the intermediate-temperature gas distribution in \S~\ref{sec:tmls}—require resolving the Field length, which can be sub-parsec in scale. In contrast, cold gas inflows through hot halos evolve on scales of hundreds of kiloparsecs. The dynamic range required to resolve both simultaneously spans over six orders of magnitude in length (or over 18 orders of magnitude in mass), making it infeasible for direct numerical resolution in large-volume simulations.

Such multi-scale problems are not new to astrophysics. A prime example is feedback mechanisms such as supernovae and AGN, which originate on sub-parsec scales but impact the evolution of entire galaxies and their surroundings. The conventional approach to handling these processes in large-scale simulations is through subgrid models, which capture the unresolved physics via effective source and sink terms -- with resounding success \citep[see, e.g., reviews by][]{Naab2016,Valentini2025}. 
 Closely related are subgrid models of star formation in the interstellar medium, which explicitly rely on an unresolved multiphase structure of the gas, typically consisting of cold, star-forming clouds embedded in a warmer or hotter ambient medium. These models translate small-scale phase structure and exchange processes into effective equations governing star formation rates, pressurization, and feedback coupling.

In this sense, classical multiphase ISM models can be viewed as early and highly influential examples of multiphase subgrid modeling. Notable implementations include the two-phase ISM model of \citet{Yepes1997}, the effective equation-of-state approach of \citet{Springel2002}, and subsequent refinements incorporating more detailed feedback and phase exchange prescriptions \citep[e.g.][]{Braun2012}. While originally developed to regulate star formation on galactic scales, these models are fundamentally based on assumptions about unresolved cold gas phases and their interaction with the surrounding medium. The multiphase subgrid models discussed below extend this philosophy beyond the star-forming ISM, aiming to capture analogous phase interactions (a) more explicitly often by modeling the different phases through a multi-fluid approach and (b) in regions beyond the ISM, specifically in the CGM and ICM, where the physical processes differ but the underlying multiscale challenge is closely related.

The need for subgrid modeling of multiphase gas flows arises from the fact that current cosmological simulations are far from resolving the observed properties of cold gas structures in the CGM and ISM. High-z quasar absorption line observations demonstrate the prevalence of small ($\lesssim 10\,$pc) cold gas structures   \citep{Rauch1999,Schaye2007,Crighton2015,Lan2017}, yet most state-of-the-art galaxy formation simulations operate at much coarser resolutions. The numerical convergence of basic cold gas properties, such as the mass of $10^4\,$K gas in the CGM, thus remains an open issue (see \S\ref{sec:ehanced_resolution_cgm}), making it difficult to disentangle physical discrepancies from numerical resolution effects.

Furthermore, while increased resolution in zoom-in simulations can improve the representation of multiphase gas, the memory and computational requirements make such approaches impractical for large statistical samples of halos. A dedicated multiphase subgrid model provides a potential solution by capturing the essential physics of unresolved cold gas phases while maintaining computational feasibility. However, akin to the feedback subgrid models mentioned above, the additional source and sink terms are relatively unconstrained, thus, one sacrifices to some extent the predictive power of the simulation. This is potentially a major issue for subgrid models of multi-phase gas flows as the majority of cells in a cosmological simulations would be directly affected (which is a major difference from the existing feedback subgrid models).

Despite the prevalence of multiphase flows in astrophysics, subgrid models for these processes are still relatively underdeveloped compared to, for example, stellar and AGN feedback models.
In contrast, in engineering applications, multiphase subgrid models are widely used to model complex fluid interactions, such as oil-water separation in petroleum engineering, water-air interactions in hydrodynamics, or gas-solid flows in combustion systems. These models are essential for accurately capturing unresolved phase interactions in turbulent flows, and their development has been extensively refined over decades \citep[e.g.,][]{Prosperetti2007}. The well-established numerical techniques in these fields offer ways forward that could potentially be adapted to astrophysical simulations.

Current efforts in astrophysical simulations have started to adopt some of these approaches, leading to the development of various multiphase subgrid models. Below, we summarize several key implementations:
\begin{enumerate}
  \item \textbf{Early two-phase models in SPH simulations:} 
   One of the earliest implementations of a multiphase subgrid model in galaxy formation simulations was by \citet{Semelin2002}, who introduced a two-phase scheme in an smoothed particle hydrodynamics (SPH) framework. Their model included a `cold' ($T < 10^4$ K) phase represented by sticky particles and a `warm' phase, with transitions between the two governed by cooling and feedback. While simplistic, this model demonstrated the feasibility of incorporating multiphase gas dynamics into galaxy simulations. Similar approaches were followed by \citet{Berczik2003} and \citet{Harfst2004}, extending the idea to three-phase representations.

   \item \textbf{Decoupling cold gas in SPH winds:}
   \citet{Scannapieco2006} introduced a `decoupled' cold phase in SPH-based models, allowing low-entropy gas to evolve separately from the hot medium. While initially motivated by the `overcooling problem’ in galactic winds \citep[e.g.,][]{Kim2015, Smith2018}, this approach can be seen as an implicit multiphase model where cold gas is dynamically distinct from its surroundings, leading to more efficient wind launching. This method shares similarities with implementations of supernova-driven winds where particles are temporarily decoupled from hydrodynamic forces \citep{Springel2002,Okamoto2005MNRAS.363.1299O,Oppenheimer2006MNRAS.373.1265O}.

 \item \textbf{Eulerian-Lagrangian multiphase models:}
   More recent efforts have shifted toward explicitly tracking multiphase structure within a single numerical element. In Eulerian-Lagrangian approaches such as those of \citet{Huang2020,Huang2021} and \citet{Smith2024}, the cold phase is represented by particles embedded within hot gas cells. These models introduce physically motivated source and sink terms between the phases, similar to subgrid turbulence models in engineering. However, their reliance on discrete particles introduces challenges in representing the cold gas velocity dispersion, particularly in turbulent environments. Thus, the application of these models has been (as above) mainly focused on galactic winds / fountain flows, where in contrast the particle representation of the colder phase allows one in principle to represent several bulk flow components within a cell.

 \item \textbf{Eulerian-Eulerian multiphase models:}
   An alternative approach is the fully Eulerian multifluid model, where the cold and hot phases are treated as distinct but interlinked fluids. Recently, \citet{Butsky2024} and \citet{Das2024} (building on the multi-fluid implementation of \citealp{Weinberger2023}) have implemented two-phase Eulerian-Eulerian subgrid models into \texttt{ENZO} and \texttt{Arepo}, respectively. Aside from the well-defined mathematical formulation of a multi-fluid Eulerian approach, an advantage of this method is that it can be applied for arbitrary volume filling fractions of the phases\footnote{This is the case if each phase is modeled as a compressible component \citep[such as in the implementation of ][]{Weinberger2023} and not assumed to be pressureless (which only applies for small volume filling fractions).}, and can represent the dispersion of the gas well -- however, each phase has only one mean velocity component, and thus, it fails to capture multiple bulk flows. 
 \end{enumerate}
 
While significant progress has been made in incorporating multiphase subgrid models into astrophysical simulations, several key challenges remain:
\begin{itemize}
\item \textit{Convergence and predictive power:} The effectiveness of a subgrid model hinges on its ability to yield converged results as resolution increases, i.e., to give the same results whether the colder phase is resolved or not. Many current models still rely on empirical tuning, reducing their predictive power when applied outside their calibration regime. To overcome this challenge, it is important to compare the subgrid models in idealized setups to fully resolved simulations of multi-phase gas flows discussed earlier in this review, and to incorporate physically motivated, well tested coupling terms between the phases with minimal free parameters.
\item \textit{Coupling with additional physics:}
  The subgrid models crucially rely on the coupling terms, and thus, on the results of small-scale multi-phase simulations. As mentioned in \S~\ref{sec:tmls}, for instance, there are still remaining uncertainties in these simulations, e.g., how viscosity and thermal conduction affect the mixing and small scale structure is still under debate.
  In addition, it should be noted that current subgrid models do not include magnetic fields self-consistently, and mostly focus on only two phases.
\item \textit{Bridging the gap between local and cosmological scales:} Even if current implementations can effectively represent multiphase gas flows in and around galaxies, there is still a long way to go before they can be fully integrated into large-volume cosmological simulations. Two crucial challenges are the need to self-consistently couple multiphase subgrid models with existing subgrid prescriptions for feedback and star formation on cosmological scales, and ensuring that these models remain computationally feasible without excessive parameter tuning. Developing a framework that can dynamically adapt to different environments—such as the ISM, CGM, and ICM -- while maintaining physical consistency across scales will be essential for future progress.

\item  \textit{Connections to observations:} Multiphase subgrid models do not follow the (unresolved) morphology of colder gas within a simulation cell. As a result, connecting these simulations to observations -- such as absorption or emission line studies, which are sensitive to the distribution and structure of the gas, is challenging and requires additional modeling or post-processing. Furthermore, intermediate-temperature gas is not modeled explicitly, thus, tracers such as OVI must be inferred or `painted on' in post-processing using approximate methods.
  \end{itemize}
  Ultimately, a robust multiphase subgrid model could be an attractive pathway for capturing the evolution of gas in and around galaxies in a converged manner, particularly in future large-scale simulations. If the challenges outlined above can be overcome, such models may play a crucial role in shaping the next generation of large scale cosmological simulations.

\subsection{Open questions and remaining challenges}

Given observational, theoretical, and computational advances, `multiphase gas dynamics' has attracted considerable interest in recent years. In this review, we have mapped out several key simulation setups and results, as well as highlighted many of the remaining challenges and open questions within each sub-field.

One common challenge that must be overcome is the variation in the diagnostic quantities analyzed -- both within individual sub-communities using different analysis methods and, even more so, between different simulation approaches. 
In other words, it remains a major open question how different simulation setups -- and the physical scales they represent -- can be connected, and how insights from each can inform one another. 
An interesting potential avenue is the use of `subgrid models' as discussed in \S~\ref{sec:subgrid}, but also increased integration with more traditional techniques, such as adaptive refinement (see \S~\ref{sec:ehanced_resolution_cgm}) or so-called `handshake simulations', which may play an important role in bridging these scales.

Another crucial area that requires further development is increased fidelity in connecting to observations, which ultimately provide the benchmark for all the theories and models discussed in this review. Of particular importance are radiative transfer models and tools for generating synthetic observations, which are essential for making meaningful, quantitative comparisons between simulations and data.

Despite the outstanding questions, the field has made tremendous progress in recent years. 
Many of the processes that govern multiphase gas flows are now better understood, and simulations are increasingly able to capture the relevant physics across a wide range of environments. 
Still, there remain many exciting challenges ahead, making this a vibrant and rapidly evolving area of astrophysics with much remaining to explore.

\bmhead{Acknowledgments}
We thank the multiphase gas community, and in particular the participants of the Multiphase Workshops 2023 and 2025, for many stimulating discussions that shaped this review. We are particularly grateful to the Jonathan Stern, Peng Oh, and the referees for valuable comments on earlier drafts.
MG thanks the Max-Planck-Society for support through the MPRG and the the European Union for support through ERC-2024-STG 101165038 (ReMMU). E.E.S. acknowledges support from NASA ATP grant 80NSSC22K0720 and the David and Lucile Packard Foundation.

\end{document}